\documentclass{aa}

\usepackage{amsmath}
\usepackage{graphicx}
\usepackage{color}
\usepackage{txfonts}
\usepackage[squaren,Gray]{SIunits}
\usepackage{natbib}
\begin{document} 
  \renewcommand{\vec}[1]{\mathbf{#1}}
  \newcommand{\diver}{\vec{\nabla} \cdot}  
  \newcommand{\divs}{\vec{\nabla}}  
  \newcommand{\grad}{\vec{\nabla}}  
  \newcommand{\vgrad}{(\vec{v} \cdot \vec{\nabla})}  
  \newcommand{\dvgrad}{(\vec{\Delta v} \cdot \vec{\nabla})}  
  \newcommand{\dvkgrad}{(\vec{\Delta v}_k \cdot \vec{\nabla})}  
  \newcommand{\vggrad}{(\vec{v_{\mathrm{g}}} \cdot \vec{\nabla})}  
  \newcommand{\vdgrad}{(\vec{v_{\mathrm{d}}} \cdot \vec{\nabla})}  
\newcommand*\Laplace{\mathop{}\!\mathbin\bigtriangleup}
   \title{Small dust grain dynamics on adaptive mesh refinement grids}

   \subtitle{I. Methods.}

   \author{U. Lebreuilly
          \inst{1}
          \and
          B. Commer\c con\inst{1}
          \and G. Laibe \inst{1}}

   \institute{\'Ecole normale sup\'erieure de Lyon, CRAL, UMR CNRS 5574, Universit\'e de Lyon , 46 All\'ee d Italie, 69364 Lyon Cedex 07, France
              \\
              \email{ugo.lebreuilly@ens-lyon.fr}      
             }

   \date{}

 
  \abstract
   {Small dust grains are essential ingredients of star, disk and planet formation.}
   {We present an Eulerian numerical approach to study small dust grain dynamics  in the context of star and protoplanetary disk formation. It is designed for finite volume codes. We use it to investigate dust dynamics during the  protostellar collapse.}
   { We present a method to solve the monofluid equations of gas and dust mixtures with several dust species in the diffusion approximation implemented in the adaptive-mesh-refinement code {\ttfamily RAMSES}.  It uses a finite volume second-order Godunov method with a predictor-corrector MUSCL scheme to estimate the fluxes between the grid cells.}
   {We benchmark our method against six distinct tests, \textsc{dustyadvect}, \textsc{dustydiffuse}, \textsc{dustyshock}, \textsc{dustywave},  \textsc{settling} and \textsc{dustycollapse}. We show that the scheme is second-order accurate in space on uniform grids and intermediate between second- and first-order on non-uniform grids. We apply our method on various \textsc{dustycollapse} simulations of $1 M_{\odot}$ cores composed of gas and dust.  }
   {We developed an efficient approach to treat gas and dust dynamics in the diffusion regime on grid-based codes. The canonical tests were successfully passed. In the context of protostellar collapse, we show that dust is less coupled to the gas in the outer regions of the collapse where grains larger than $\simeq 100~ \micro\meter$ fall significantly faster than the gas.}


   \maketitle
%

\section{Introduction}

Dust grains are thought to represent on average only $1\% $ of the mass of the diffuse interstellar medium  \citep[ISM,][]{1977ApJ...217..425M,2001ApJ...548..296W}. Their typical size distribution is usually modeled as a power law \citep[][hereafter MRN]{1977ApJ...217..425M}. However, recent works indicate that this is probably not the case in the dense parts of the ISM \citep{2010Sci...329.1622P,2011A&A...525A.103C,2018NatAs.tmp...80H}. In addition, a dynamical size sorting may operate in molecular clouds \citep{2016MNRAS.456.4174H,molecul} or during the protostellar collapse \citep{2017MNRAS.465.1089B}. This sorting leads to important variations in the dust-to-gas ratio, especially for large dust grains. 

The dust budget plays an essential role during the star and disk formation processes. On the one hand, it regulates the thermal budget of the medium through its blackbody emission, its absorption, and the photoelectric effect. On the other hand, the charge on the grain surface controls, through their number, the ambipolar, ohmic and hall resistivities during the protostellar collapse \citep{2016A&A...592A..18M}. \cite{2016MNRAS.460.2050Z} show that removing grains smaller than $\sim 100~\nano\meter$  from the dust distribution enhances the effects of ambipolar diffusion, triggering the formation of a rotationally supported disk. As we need to know the dust concentration, a treatment of its dynamics is therefore essential to improve our understanding of stellar, disk, and planet formation.


  Bifluid models of strongly coupled dust and gas mixtures are difficult to integrate numerically. For Lagrangian-based methods they tend to produce spurious dust aggregates when the grains are accumulated below the resolution length of the gas \citep{2012MNRAS.423.1450A}. They also require extremely high spatial resolution in order to capture correctly the dephasing between gas and dust, i.e., the dissipation of energy through the drag \citep{2012MNRAS.420.2365L}. Some efficient bifluid treatments were developed to cope with these issues  \citep[e.g., semi implicit methods,][]{2014MNRAS.443..927L,2015MNRAS.454.4114L,2017ARep...61.1044S,2018ARep...62..455S}.
As an alternative way to model correctly these mixtures, \cite{2014MNRAS.440.2136L} proposed an algorithm for smoothed Particle Hydrodynamic codes \citep[SPH,][]{1977AJ.....82.1013L,1977MNRAS.181..375G} using a monofluid formalism. It has been proven to be efficient in the so-called diffusion approximation valid for strongly coupled mixtures \citep{2015MNRAS.454.2320P}.  More recently, \cite{2018MNRAS.476.2186H} improved this method and adapted it to treat simultaneously several dust species.  \cite{2017ApJ...849..129L} and \cite{2018MNRAS.478.2737C} have applied the monofluid formalism on a cylindrical grid adapted to protoplanetary disk simulations with an Eulerian approach.

Adaptive Mesh Refinement  \citep[AMR,][]{1984JCoPh..53..484B}  codes are a complementary approach to SPH in astrophysical or cosmological simulations.  \cite{2008A&A...482..371C} show in the context of protostellar collapse that the two methods lead to converged results, but AMR codes capture details more efficiently thanks to the versatility of the refinement criteria. The adaptive mesh is a high performance tool for star formation simulations where a wide range of physical and dynamical scales are considered \citep[e.g.,][]{2018A&A...611A..24H,2018arXiv180108193V}. In this paper, we present a numerical Eulerian approach of dust dynamics in the diffusion approximation on AMR grids and its implementation in the {\ttfamily RAMSES} code \citep{2002A&A...385..337T},  as well as an application to the first collapse \citep{1969MNRAS.145..271L}.

The paper is organized as follows. In Sect. \ref{sec:theory}, we recall the theory of gas and dust mixtures and the diffusion approximation. Then the dust diffusion algorithm implemented in {\ttfamily RAMSES}  is presented in Sect. \ref{sec:numerical} followed by the validation tests in Sect. \ref{sec:tests}. The application on the dynamics of dust during the first collapse in the spherical case is shown in Sect. \ref{sec:collapse}. In Sect. \ref{sec:conclusion}, we present our conclusions and perspectives. A companion paper, paper II (Lebreuilly et al. in prep.), will discuss in detail the dust dynamics during the first collapse with initial solid-body rotation and turbulence, and with multiple grains species. \newline \newline

\section{Gas and dust mixtures in the diffusion approximation.}
\label{sec:theory}
\subsection{Grain dynamics}
First, we examine the fate of a single  grain within a gas of density $\rho_{\mathrm{g}}$ . Let us consider a grain of  radius $s_{\mathrm{grain}}$, mass $m_{\mathrm{grain}}$, and intrinsic density $\rho_{\mathrm{grain}}$. It exchanges momentum with the surrounding gas molecules through microscopic collisions. Macroscopically, this leads to a drag force on the grain that writes
  \begin{eqnarray}
    \vec{F} = -\frac{m_{\mathrm{grain}}}{t_{\mathrm{s,grain}}}  \vec{\Delta v},
  \end{eqnarray}
  where the stopping time  of the grain $t_{\mathrm{s,grain}}$, is the typical time required by the dust grain to adjust its dynamics to a change in the gas velocity, and $ \vec{\Delta v}$ is the differential velocity between the grain and the gas. 
    In astrophysical regimes the mean free path of the gas is much larger than the size of the grain. In this case the grain stopping time is given by \citep{1924PhRv...23..710E} 
   \begin{eqnarray}
   \label{eq:tstop1}
   t_{\mathrm{s,grain}} \equiv \sqrt{\frac{\pi \gamma}{8}} \frac{\rho_{\mathrm{grain}}}{\rho_{\mathrm{g}}}\frac{s_{\mathrm{grain}}}{c_{\mathrm{s}}} ,
 \end{eqnarray}
where $c_{\mathrm{s}}$ and $\gamma$  denote the sound speed and the adiabatic index of the gas, respectively.  This is a particular case of a linear drag regime.

 Large and dense grains are less coupled  to the gas. If $c_{\mathrm{s}}$ and $\rho_{\mathrm{g}}$ increase, the coupling of the two phases becomes more important as  the collision rate between gas and dust particles increases. We note that the opposite drag force applies on the gas particles for momentum conservation.

  \subsection{The monofluid formalism}
    \label{subsec:The monofluid formalism} 
    
    \subsubsection{Gas and dust bifluids}
  A usual model for gas and dust mixtures consists of two fluids interacting via the drag term \citep{1962JFM....13..120S}.  Due to the momentum exchanges between grains and gas particles, collective effects of the dust grains play an important role in the mixture dynamics. A continuous fluid description of the  dust is more practical than treating the grains independently although this description might be inappropriate for very large particles.  The dust fluid is characterized by a density $\rho_{\mathrm{d}}$ and an advection velocity $\vec{v_{\mathrm{d}}}$. The dust density $\rho_{\mathrm{d}}$ must not be confused with the intrinsic grain density $\rho_{\mathrm{grain}}$. This fluid is considered pressureless and inviscid since the collisions with other grains are much less frequent than with gas particles. Through the drag, dust feels the forces exerted on the gas. The surrounding gas is  described, as usual, as a fluid with a density $\rho_{\mathrm{g}} $ and an advection velocity $\vec{v_{\mathrm{g}}}$.  
 
The bifluid formulation of gas and dust  mass and momentum conservation writes
     \begin{eqnarray}
     \label{eq:bifl}
      \frac{\mathrm{d}_{\mathrm{g}} \rho_{\mathrm{g}}}{\mathrm{d} t} & = & - \rho_{\mathrm{g}} (\diver\vec v_{\mathrm{g}}), \nonumber
       \\
         \frac{\mathrm{d}_{\mathrm{d}} \rho_{\mathrm{d}}}{\mathrm{d} t} & = & - \rho_{\mathrm{d}} (\diver\vec v_{\mathrm{d}}), \nonumber
       \\  
        \rho_{\mathrm{g}} \frac{\mathrm{d_{\mathrm{g}}} \vec{v}_{\mathrm{g}}}{\mathrm{d} t} &= & -\grad P_{\mathrm{g}}+ \rho_{\mathrm{g}} \vec{f}_{\mathrm{g}} + \rho_{\mathrm{g}} \vec{f} +  K  \vec{\Delta v},\nonumber  \\
      \rho_{\mathrm{d}} \frac{\mathrm{d_{\mathrm{d}}} \vec{v}_{\mathrm{d}}}{\mathrm{d} t} &= & \rho_{\mathrm{d}} \vec{f}_{\mathrm{d}}  + \rho_{\mathrm{d}} \vec{f}  -K  \vec{\Delta v},
      \nonumber\\
   {\frac{\mathrm{d}_{\mathrm{g}} e_{\mathrm{g}}}{\mathrm{d} t} }& {=}&{    - \frac{P_{\mathrm{g}}}{\rho_{\mathrm{g}}}\nabla \cdot  \vec{v}_{\mathrm{g}} + \frac{K}{\rho_{\mathrm{g}}} \vec{\Delta v}\cdot \vec{\Delta v} },
  \end{eqnarray} 
  where $ \vec{\Delta v} \equiv  \vec{v_{\mathrm{d}}} -\vec{v_{\mathrm{g}}}$ is the differential velocity between the two fluids;  $\vec{f}$ is the  specific force acting on both gas and dust; and $\vec{f_{\mathrm{g}}}$ and $\vec{f_{\mathrm{d}}}$ are the  specific forces, drag and pressure force excluded, affecting the gas or the dust, respectively.  In the rest of the article, we assume the hydrodynamical case where $\vec{f_{\mathrm{g}}}$ and $\vec{f_{\mathrm{d}}}$ are zero, and $\vec{f}$ is either zero or the gravity force.   The gas pressure $P_{\mathrm{g}}$ is given by $P_{\mathrm{g}}= (\gamma -1 ) \rho_{\mathrm{g}}e_{\mathrm{g}}$, $e_{\mathrm{g}}$ being the gas specific  internal energy.  The gas derivative  $\frac{\mathrm{d}_{\mathrm{g}}}{\mathrm{d} t} \equiv \frac{\partial }{\partial t} + \left(\vec{v}_{\mathrm{g}} \cdot \nabla\right)  $ differs from the dust derivative $\frac{\mathrm{d}_{\mathrm{d}}}{\mathrm{d} t} \equiv \frac{\partial }{\partial t} + \left(\vec{v}_{\mathrm{d}} \cdot \nabla\right) $. 
   We also define the stopping time as
    \begin{eqnarray}
   \label{eq:tstop}
   t_{\mathrm{s}} \equiv \sqrt{\frac{\pi \gamma}{8}} \frac{\rho_{\mathrm{grain}}}{\rho}\frac{s_{\mathrm{grain}}}{c_{\mathrm{s}}} = \frac{\rho_{\mathrm{g}}}{\rho} t_{\mathrm{s,grain}} ,
 \end{eqnarray}
   and the drag coefficient $K$  as
  \begin{eqnarray}
   K \equiv \frac{\rho_{\mathrm{d}} \rho_{\mathrm{g}}}{\rho  t_{\mathrm{s}}}.
  \end{eqnarray}
   \subsubsection{Two-phases mixture}
  
  Alternatively, it is possible to model the gas and dust mixture as a single fluid made of two phases of total density 
$ \rho \equiv  \rho_{\mathrm{d}}+\rho_{\mathrm{g}}$ \citep{2014MNRAS.440.2136L,2014MNRAS.440.2147L}. It is moving at the mixture barycentric velocity $\vec{v}$:
    \begin{eqnarray*}    
\vec{v}  \equiv \frac{\rho_{\mathrm{g}} \vec{v_{\mathrm{g}}} + \rho_{\mathrm{d}} \vec{v_{\mathrm{d}}}}{\rho_{\mathrm{g}}+\rho_{\mathrm{d}}}.
  \end{eqnarray*}
     We also define the dust ratio 
  \begin{eqnarray*}
  \epsilon \equiv \frac{\rho_{\mathrm{d}}}{\rho},
  \end{eqnarray*} 
 which must not be confused with the dust-to-gas ratio $ \frac{\rho_{\mathrm{d}}}{\rho_{\mathrm{g}}}$. 
  Using these new quantities, it is straightforward to write 
     \begin{eqnarray*}
  \vec{v_{\mathrm{g}}}  = \vec{v} - \epsilon \vec{\Delta v},
  \end{eqnarray*}
  \begin{eqnarray*}
  \vec{v_{\mathrm{d}}}  =  \vec{v} + (1-\epsilon)\vec{\Delta v},
  \end{eqnarray*}
 \begin{eqnarray*}
  {\rho_{\mathrm{g}}} = \left(1-\epsilon \right) \rho,
 \end{eqnarray*}
  \begin{eqnarray*}
  {\rho_{\mathrm{d}}} = \epsilon \rho.
 \end{eqnarray*}
  After some developments, we write the conservation of mass and momentum for the mixture and of the  gas internal energy 
     \begin{eqnarray}    
     \label{eq:fullmono}
      \frac{\mathrm{d} \rho}{\mathrm{d} t} & = & -\rho (\diver\vec v), \nonumber
       \\   \frac{\mathrm{d} \vec{v}}{\mathrm{d} t} &= &  -\frac{\grad P_{\mathrm{g}}}{ \rho}+ \vec{f} - \frac{1}{\rho} \diver \left( \epsilon (1-\epsilon) \rho \vec{\Delta v} \otimes \vec{\Delta v} \right), \nonumber\\ \frac{\mathrm{d} \epsilon}{\mathrm{d} t}  &=& -\frac{1}{\rho}\diver \left(\epsilon (1-\epsilon)\rho \vec{\Delta v}\right), \nonumber\\ \label{eq:fullmono2}\frac{\mathrm{d} \vec{\Delta v}}{\mathrm{d} t} &=& \frac{\grad P_{\mathrm{g}}}{(1-\epsilon) \rho} -\frac{\vec{\Delta v}}{t_{\mathrm{s}}}   - \dvgrad \vec{v} + { \frac{1}{2} \nabla\left((2 \epsilon -1)\vec{\Delta v} \cdot \vec{\Delta v} \right) }\nonumber\\
       & & {  +  (1-\epsilon) \vec{\Delta v} \times  (\nabla \times (1-\epsilon) \vec{\Delta v}  )  -\epsilon \vec{\Delta v} \times  (\nabla \times  \epsilon \vec{\Delta v}  )},\nonumber\\
        { \frac{\mathrm{d} e_{\mathrm{g}}}{\mathrm{d} t}} & { =}& {  -\frac{P_{\mathrm{g}}}{\rho (1-\epsilon)}\nabla \cdot \left( \vec{v} - \epsilon \Delta\vec{v} \right) + (\epsilon \Delta\vec{v}\cdot \nabla) e_{\mathrm{g}} +\epsilon \frac{\vec{\Delta v}\cdot \vec{\Delta v}}{t_{\mathrm{s}}}}.
  \end{eqnarray} 
 Here only one Lagrangian derivative $\frac{\mathrm{d}}{\mathrm{d} t} \equiv \frac{\partial }{\partial t} + \left(\vec{v} \cdot \nabla \right) $ has to be defined. This system must be closed by the gas equation of state. We note the presence of the last terms in the differential velocity equation that were omitted in previous studies.
  
At this point, the monofluid approach is a dual reformulation of the dust and gas fluid Eqs. (\ref{eq:bifl}). However, it may be much more convenient for numerical simulations since it requires only one resolution scale (the mixture resolution) and it avoids the artificial trapping of dust particles  \citep{2014MNRAS.440.2136L}.
 
 \subsubsection{Diffusion approximation}

 \cite{2014MNRAS.440.2136L}  show that, for a strong drag regime, i.e.,\ when the stopping time is short compared with the dynamical timescale $t_{\mathrm{dyn}}$ of the system, two simplifications can be made.

 In the so-called diffusion approximation,  $\left|\left|\vec{\Delta v}^2/\vec{ v}^2\right|\right|$, $ \left|\left|\vec{\Delta v}\otimes \vec{\Delta v}/\vec{ v}^2\right|\right| $,  $\left|\left| \vec{\Delta v} \times  (\nabla \times (1-\epsilon) \vec{\Delta v}  )/\vec{ v}^2\right|\right|$ and $\left|\left| \vec{\Delta v} \times  (\nabla \times \epsilon \vec{\Delta v}  )/\vec{ v}^2\right|\right|$ are second-order in $\mathrm{St}\equiv \left(t_{\mathrm{s}}/t_{\mathrm{dyn}} \right)$, where $\mathrm{St}$ is the Stokes number.  \footnote{ $\left|\left|.\right|\right|$ indicates the L2 norm of either a tensor or a vector.} 
 System (\ref{eq:fullmono}) reduces to
\begin{eqnarray} 
  \frac{\mathrm{d} \rho}{\mathrm{d} t}  & = & -\rho (\diver\vec v),  \nonumber  \\  \frac{\mathrm{d} \vec{v}}{\mathrm{d} t} &= & -\frac{\grad P_{\mathrm{g}}}{ \rho}+\vec{f},\nonumber  \\ \frac{\mathrm{d} \epsilon}{\mathrm{d} t}  &=& -\frac{1}{\rho}\diver \left(\rho \epsilon (\vec{\Delta v}- \epsilon\vec{\Delta v})\right), \nonumber \\
  \frac{\mathrm{d} \vec{\Delta v}}{\mathrm{d} t} &=& \frac{\grad P_{\mathrm{g}}}{(1-\epsilon) \rho} -\frac{\vec{\Delta v}}{t_{\mathrm{s}}}   - \dvgrad \vec{v}, \nonumber\\
     {    \frac{\mathrm{d} e_{\mathrm{g}}}{\mathrm{d} t} }& { =}& {  -\frac{P_{\mathrm{g}}}{\rho (1-\epsilon)}\nabla \cdot  \vec{v}  + \left(\epsilon \Delta\vec{v}\cdot \nabla\right) e_{\mathrm{g}}}.
  \end{eqnarray}
   The errors caused by the diffusion approximation are marginal as long as $\mathrm{St} \ll 1$.  We note that the first-order term $ -\frac{P_{\mathrm{g}}}{\rho (1-\epsilon)}\nabla \cdot \left(\epsilon \Delta\vec{v} \right) $ in the energy equation also simplifies owing to energy conservation, as explained by \cite{2015MNRAS.454.2320P} in their second footnote. Another approximation known as the terminal velocity approximation \citep{2005ApJ...620..459Y,2008ApJ...675.1549C} can be made when $ \mathrm{St} \ll 1 $. In this case $\left|\left|\frac{\mathrm{d}\vec{\Delta v}}{\mathrm{d} t}/\left(\frac{\vec{\Delta v}}{t_{\mathrm{s}}}\right)\right|\right|$ and $\left|\left|\dvgrad \vec{v} /\left(\frac{\vec{\Delta v}}{t_{\mathrm{s}}}\right)\right|\right|$ are transitory terms, and are thus neglected.   
A consequence of the terminal velocity approximation for linear drag regimes is that the differential velocity depends directly on the force balance
 \begin{eqnarray*}
 \vec{\Delta v}=  t_{\mathrm{s}}\frac{\grad P_{\mathrm{g}}}{(1-\epsilon) \rho}. 
 \end{eqnarray*}
 
 It should be noted that this expression can change if other forces apply on the dust or the gas, e.g., the radiation or the Lorentz forces.  In the terminal velocity approximation, we finally obtain
  \begin{eqnarray} 
  \label{eq:simpl}
  \frac{\mathrm{d} \rho}{\mathrm{d} t}  & = & -\rho (\diver\vec v), \nonumber  \\  \frac{\mathrm{d} \vec{v}}{\mathrm{d} t} &= & - \frac{\grad P_{\mathrm{g}}}{\rho}+\vec{f},\nonumber \\ \frac{\mathrm{d} \epsilon}{\mathrm{d}t} &=& -\frac{1}{\rho}\diver \left(\epsilon t_{\mathrm{s}} \grad{P_{\mathrm{g}}} \right), \nonumber \\  {  \frac{\mathrm{d} e_{\mathrm{g}}}{\mathrm{d} t} }& { =}& {  -\frac{P_{\mathrm{g}}}{\rho (1-\epsilon)}\nabla \cdot  \vec{v}\nonumber + \left(\epsilon t_{\mathrm{s}}\frac{\grad P_{\mathrm{g}}}{(1-\epsilon) \rho}\cdot \nabla\right) e_{\mathrm{g}}.}
 \end{eqnarray}

  The two first equations are identical to pure hydrodynamics when only gas is involved. The third equation describes the evolution of the dust ratio. Using a conservative formulation, it can be expressed as an advection equation 
  \begin{eqnarray}
 \frac{\partial   \rho \epsilon}{\partial t} +\diver \left[ \rho \epsilon \left( \vec{v} +  \frac{ t_{\mathrm{s}} \grad{P_{\mathrm{g}}}}{ \rho}\right) \right] = 0. \label{eq:eqdiff}
  \end{eqnarray}
   In this formulation, the equation is almost identical to mass conservation but with a different advection velocity due to the dephasing between the dust and  the barycenter. 
 The specific internal energy equation is similar to pure hydrodynamics with an additional term  that accounts for the back-reaction of dust on the gas.

 \subsubsection{($\mathcal{N}+1$) phase mixtures}
 In the diffuse interstellar medium a grain size distribution commonly modeled as a power law \citep{1977ApJ...217..425M} is observed. It is therefore more realistic to consider several dust species in order to reproduce the observed dust size distribution. 
 
 In this perspective, the previous monofluid formalism has been extended  to $\left(\mathcal{N}+1\right)$ phase mixtures with $\mathcal{N}$ distinct dust species and a gas phase \citep{2014MNRAS.444.1940L,2018MNRAS.476.2186H}. They show that, in the diffusion approximation the monofluid set of equations writes
  \begin{eqnarray} 
  \label{eq:simplm}
  \frac{\mathrm{d} \rho}{\mathrm{d} t}  & = & -\rho (\diver\vec v), \nonumber  \\  \frac{\mathrm{d} \vec{v}}{\mathrm{d} t} &= & - \frac{\grad P_{\mathrm{g}}}{\rho}+\vec{f},\nonumber \\ \frac{\mathrm{d} \epsilon_k}{\mathrm{d}t} &=& -\frac{1}{\rho}\diver \left(\epsilon_k T_{\mathrm{s},k} \grad{P_{\mathrm{g}}} \right),\ \forall k \in \left[1,\mathcal{N}\right], \nonumber \\   { \frac{\mathrm{d} e_{\mathrm{g}}}{\mathrm{d} t}} & { =}& {  -\frac{P_{\mathrm{g}}}{\rho (1-\mathcal{E})}\nabla \cdot  \vec{v}   + \left(\mathcal{E} \mathcal{T}_{\mathrm{s}}\frac{\grad P_{\mathrm{g}}}{(1-\mathcal{E}) \rho}\cdot \nabla\right) e_{\mathrm{g}}},
 \end{eqnarray}
    where  $\epsilon_k$ is the dust ratio of the phase $k$ and $T_{\mathrm{s},k} $  is the effective stopping time of the dust phase $k$  defined as
\begin{eqnarray*}T_{\mathrm{s},k} \equiv \frac{t_{\mathrm{s},k}}{1-\epsilon_k}- \sum_{l=1}^{\mathcal{N}} \frac{\epsilon_l}{1-\epsilon_l} t_{\mathrm{s},l},
\end{eqnarray*}
where $t_{\mathrm{s},k} $ is the individual stopping time of the phase $k$. The $\sum_{l=1}^{\mathcal{N}} \frac{\epsilon_l}{1-\epsilon_l} t_{\mathrm{s},l}$  term  accounts for the interaction between dust species that is due to their cumulative back-reaction on the gas. 
 We also introduces the total dust ratio 
  \begin{eqnarray*}
 {  \mathcal{E} \equiv \sum_{l=1}^{\mathcal{N}} \epsilon_l,}
 \end{eqnarray*}
 and the mean stopping time
   \begin{eqnarray*}
  { \mathcal{T}_{\mathrm{s}} \equiv \frac{1}{\mathcal{E}} \sum_{l=1}^{\mathcal{N}} \epsilon_l T_{\mathrm{s},l}.}
 \end{eqnarray*}
 The gas and dust densities are simply given by
 \begin{eqnarray*}
  \label{eq:epsmult}
 \rho_{\mathrm{g}}&\equiv& \left(1-\mathcal{E}\right) \rho,\nonumber \\
 \rho_{\mathrm{d},k}& \equiv& \epsilon_k \rho.
 \end{eqnarray*}
When $\mathcal{N}=1$, Eqs. (\ref{eq:simplm}) reduce to the formulation for a single dust phase.

Combining the different equations with mass conservation in the Lagrangian frame, co-moving with the barycenter leads to the formulation of the system of equations in a conservative form
  \begin{eqnarray} 
  \label{eq:simpl_cons}
  \frac{\partial \rho}{\partial t}  + \diver \left[ \rho \vec v \right]&=&0, \nonumber  \\  \frac{\partial \rho \vec{v}}{\partial t} + \diver \left[P_{\mathrm{g}} \mathbb{I} + \rho (\vec{v}\otimes  \vec{v}) \right] &=&\rho \vec{f},\nonumber \\ \frac{\partial   \rho_{\mathrm{d},k} }{\partial t} +\diver \left[ \rho_{\mathrm{d},k} \left( \vec{v} +  \frac{ T_{\mathrm{s},k} \grad{P_{\mathrm{g}}}}{ \rho}\right) \right] &=& 0 , \nonumber \\
  \frac{\partial E}{\partial t} + \diver \left[  (E+P_{\mathrm{g}})\vec{v}\right]&=&\nabla \cdot \left[ \frac{\mathcal{E} \mathcal{T}_{\mathrm{s}}}{1-\mathcal{E}} \frac{\grad{P_{\mathrm{g}}}}{\rho}\frac{P_{\mathrm{g}}}{\gamma-1}\right],
 \end{eqnarray}
where  $\rho_{\mathrm{d},k} $ is  the  density of the dust phase $k$, $E \equiv \frac{1}{2} \rho \vec{v}^2 + \rho (1-\mathcal{E}) e_{\mathrm{g}} $ is the total energy of the mixture and $\mathbb{I}$ is the identity matrix.

\section{Numerical methods}
\label{sec:numerical}

\subsection{{\ttfamily RAMSES}}
\subsubsection{Basic features}
Before we present our method, we recall the main features of the  {\ttfamily RAMSES} code \citep{2002A&A...385..337T} in order to facilitate the understanding of our implementation of dust dynamics in the diffusion approximation.

 {\ttfamily RAMSES} is a finite volume Eulerian code that integrates the equation of hydrodynamics in their conservative form on an AMR grid \citep{1984JCoPh..53..484B}. Among  others applications, it has been extended to magnetohydrodynamics \citep{2006JCoPh.218...44T,2006A&A...457..371F,2012ApJS..201...24M,Hall-pierre}, radiation hydrodynamics \citep{2011A&A...529A..35C,2013MNRAS.436.2188R,2014A&A...563A..11C,2015MNRAS.449.4380R,2015A&A...578A..12G}, and cosmic ray and anisotropic heat conduction \citep{2016A&A...585A.138D} .

For simplicity,  the hydrodynamical version of {\ttfamily RAMSES}  is presented here. It solves the Euler equations of a pure gas in their hyperbolic form
	\begin{eqnarray}
	\label{eq:god}
	\frac{\partial \mathbb{U}}{\partial t} +\diver \mathbb{F} (\mathbb{U})=0,
	\end{eqnarray}
	the state $\mathbb{U}$ vector and flux $ \mathbb{F}$ vector being 
	 \begin{eqnarray*}
	\mathbb{U}&\equiv & \left(\rho_{\mathrm{g}}, \rho_{\mathrm{g}} \vec{v}_{\mathrm{g}}, E_{\mathrm{g}}\right), \\ \mathbb{F}(\mathbb{U})&\equiv &\left(\rho_{\mathrm{g}} \vec{v}_{\mathrm{g}} , \rho_{\mathrm{g}} \vec{v}_{\mathrm{g}}\otimes\vec{v}_{\mathrm{g}}+P_{\mathrm{g}} \mathbb{I}, \vec{v}_{\mathrm{g}}(E_{\mathrm{g}}+P_{\mathrm{g}})\right),	
	\end{eqnarray*} 
	where $E_{\mathrm{g}} \equiv \frac{1}{2} \rho_{\mathrm{g}} \vec{v_{\mathrm{g}}}^2 + \rho_{\mathrm{g}} {e}_{\mathrm{g}}$ is the total energy of the gas.

\subsubsection{Godunov scheme}
	  
 Finite volume methods are based on the estimation of the average of $\mathbb{U}$ over the cells. For a 1D problem integrated over time (for $t \in [t_1,t_2]$) and space (for $x \in [x_1,x_2]$), the previous system writes 
\begin{eqnarray}
	\label{eq:godint}
  \int_{x_1}^{x_2} \left(\mathbb{U}(t_2)-\mathbb{U}(t_1)\right)\mathrm{d}x+  \int_{t_1}^{t_2} \left(\mathbb{F}(x_2)-\mathbb{F}(x_1)\right)\mathrm{d}t =0.
	\end{eqnarray}
	
Hence, the evolution of the state vector is constrained by the fluxes at the interfaces of the cells. {\ttfamily RAMSES} uses a second-order predictor-corrector Godunov method \citep{godunov} to update its value.  

At the timestep $n$ for a cell $i$, any physical quantity $A$  is discretized as  
 \begin{eqnarray*}
  A (x,t) \rightarrow A^n_i.
  \end{eqnarray*}
  Cell interfaces are denoted with half integer  subscripts, e.g, $i+1/2$ is to the interface between the cell $i$ and $i+1$. Similarly, half timesteps are also denoted with half integer   superscripts, e.g, $n+1/2$. 
  
  For a cell of length $\Delta x$ and a timestep $\Delta t$, the scheme writes
 \begin{eqnarray}
 \label{eq:Godunov}
 \mathbb{U}^{n+1}_i = \mathbb{U}^n_i - \left(\mathbb{F}^{n+1/2}_{i+1/2}-\mathbb{F}^{n+1/2}_{i-1/2}\right) \frac{\Delta t}{\Delta x},
 \end{eqnarray}
 where the discretized fluxes $\mathbb{F}^{n+1/2}_{i\pm1/2}$ ,which are the solutions of the Riemann problem at the interfaces, are approximated with a MUSCL predictor-corrector scheme \citep{1979JCoPh..32..101V} that ensures the second-order accuracy in space and in time for the Godunov method. We note that, in {\ttfamily RAMSES} $\Delta x =\Delta y =\Delta z$.
\subsubsection{AMR grid and adaptive timestep}

	The AMR grid  enables an accurate description of the regions of interest in the simulation box. Cells are tagged with a refinement level $\ell$ with a value between $\ell_{\mathrm{min}}$ for the coarsest cells and $\ell_{\mathrm{max}}$ for the finest. The length of a cell of level $\ell$ is
	  \begin{eqnarray*}	   
\Delta x_{\ell} = \frac{L_{\mathrm{box}}}{2^{\ell}},
\end{eqnarray*}
where $L_{\mathrm{box}}$ is the size of the simulation box. For a uniform grid with a unique level $\ell$, the effective resolution is simply $2^{\ell}$.

 {\ttfamily RAMSES} uses an adaptive timestep to reduce the computation time while maintaining a good accuracy for the refined cells. When a level $\ell$ is updated with a timestep $\Delta t_{\ell}$, the level $\ell+1$ is updated twice with two timesteps, $\Delta t_{1,\ell+1}$ and $\Delta t_{2,\ell+1}$, each  verifying the Courant–Friedrichs–Lewy condition \citep[][hereafter CFL]{1928MatAn.100...32C}. The two levels are synchronized with the following condition  
\begin{eqnarray*}
\Delta t_{\ell}=\Delta t_{1,\ell+1}+\Delta t_{2,\ell+1}.
\end{eqnarray*}

The state vector is updated from fine to coarse levels. For a cell of level $\ell$, three types of interfaces are possible
\begin{itemize}
\item The fine-to-coarse interfaces, when the neighbor cell is at level $\ell-1$;
\item the fine-fine interfaces, when the neighbor cell is  also at level $\ell$;
\item the coarse-to-fine interfaces, when the neighbor cell is at level $\ell+1$.
\end{itemize}  
 In {\ttfamily RAMSES}, the levels are updated from fine to coarse. During the update of level $\ell+1$, the fluxes at the interface with level $\ell$ are used to update the level $\ell+1$ and stored for the later update of the level $\ell$. During the update of the level, the state vector of the coarser levels is held constant. The fine-to-coarse and fine-fine fluxes are computed during the update of the level $\ell$. The coarse neighbor cells at level $\ell-1$ are virtually refined to compute the flux at the boundaries with level $\ell$.

\subsection{Dust diffusion scheme}

\subsubsection{Operator splitting}

In the previous section, we explain how {\ttfamily RAMSES} solves the equations of hydrodynamics in their conservative form. The monofluid formalism in the diffusion approximation has the same structure for the mass, momentum and energy conservation equation.  A total of $\mathcal{N}$ additional equations are required to follow the evolution of the dust ratios. It is still possible to write the system in a hyperbolic form, but we choose to decompose the flux in two distinct terms. The system now writes
\begin{eqnarray}
 \frac{\partial \mathbb{U}}{\partial t} +\diver \mathbb{F}_{\mathrm{H}} (\mathbb{U})+\diver \mathbb{F}_{\mathrm{\Delta}} (\mathbb{U})= 0.
\end{eqnarray}
 The new state vector  $ \mathbb{U}$ writes 
\begin{eqnarray*}
\mathbb{U}\equiv\left(\rho, \rho \vec{v}, E,  \rho_{\mathrm{d},k} \right);
\end{eqnarray*}
  the flux $\mathbb{F}_{\mathrm{H}}$ is similar to the flux of pure hydrodynamics and is given by 
\begin{eqnarray*}
  \mathbb{F}_{\mathrm{H}}(\mathbb{U}) \equiv\left(\rho \vec{v} , \rho \vec{v}\otimes\vec{v}+P_{\mathrm{g}} \mathbb{I}, \vec{v}(E+P_{\mathrm{g}}), \rho_{\mathrm{d},k}\vec{v} \right);
\end{eqnarray*}
 the flux $\mathbb{F}_{\mathrm{\Delta}}$ accounts for dust diffusion and is given by

\begin{eqnarray*}
  \mathbb{F}_{\mathrm{\Delta}}(\mathbb{U}) \equiv\left(0,0, \frac{P_{\mathrm{g}}}{\gamma -1} \vec{w_{\mathrm{g,b}}}, \rho_{\mathrm{d},k}   \vec{w_k} \right),
\end{eqnarray*}
where $\vec{w_k}$,  the differential  advection  velocity between the dust species k and the barycenter,   is
 \begin{eqnarray}
 \vec{w_k} \equiv \frac {T_{\mathrm{s},k}\grad P_{\mathrm{g}}}{\rho};
  \end{eqnarray}
   and $\vec{w_{\mathrm{g,b}}}$ is the differential velocity between the gas and the barycenter that writes
\begin{eqnarray}
 \vec{w_{\mathrm{g,b}}}=-\frac{\mathcal{E} \mathcal{T}_{\mathrm{s}}}{1-\mathcal{E}} \frac{\grad{P_{\mathrm{g}}}}{\rho}
\end{eqnarray}
 or 
\begin{eqnarray}
 \vec{w_{\mathrm{g,b}}}=- \sum_k \frac{\rho_{\mathrm{d},k}}{\rho-\sum_j\rho_{\mathrm{d},j}}\vec{w_k}.
 \label{eq:wbackk}
\end{eqnarray}
  The classical second-order Godunov method presented above is used to update the state vector. An operator splitting method is performed to solve the system in two steps. In the so-called hydrodynamical step, we only compute $\mathbb{F}_{\mathrm{H}}$ and update the state vector accordingly. An important difference  between this step and the classical hydrodynamical version  of {\ttfamily RAMSES} is that the fluid density $\rho$ is different from the gas density. As a consequence, the pressure and the wave speeds must be computed using $\rho_{\mathrm{g}}= \rho (1-\mathcal{E})$ instead of $\rho$.

\subsubsection{MUSCL scheme for dust diffusion/advection}
  The second step of the operator splitting consists in taking into account the  second flux vector $\mathbb{F}_{\mathrm{\Delta}}$.  For simplicity, we now focus on the update of the dust density  keeping in mind that the dust density and energy updates are done simultaneously. In this perspective, the following equation is solved
    \begin{eqnarray} 
 \frac{\partial \tilde{\rho}_{\mathrm{d},k}}{\partial t}=-\diver \left[\tilde{\rho}_{\mathrm{d},k} {\vec{w}_k}\right],
 \label{eq:dust}
  \end{eqnarray}
 where $\tilde{\rho}_{\mathrm{d},k}$ refers to ${\rho}_{\mathrm{d},k}$ after its advection as a passive scalar at the velocity $\vec{v}$. 
  In the remainder of the section, we omit this symbol and the $k$ index. 
    A MUSCL predictor-corrector scheme is used to compute the dust diffusion fluxes $\mathcal{F} (\rho_{\mathrm{d}}) \equiv \rho_{\mathrm{d}} \vec{w}$. The flux storage and the Godunov update are performed as they are the hydrodynamical step.
    
    Apart from the dust densities and the energy of the mixture, the variables are constant during this step. In particular, the total density $\rho$ is constant.
  \paragraph{Differential advection velocity:}
   During the diffusion step, we aim to compute the differential advection velocity ${w}_{i,j,k}^{n}$.  To get its value, the pressure gradient is evaluated at the center of the cell $i,j,k$ of length  $\Delta x$. For example, its component in the $x$-direction, is 
  \begin{eqnarray}
  \grad_{x} {P_{\mathrm{g}}}_{i,j,k}^n = \frac{{P_{\mathrm{g}}}_{i+1,j,k}^n-{P_{\mathrm{g}}}_{i-1,j,k}^n}{\Delta x_{i+1,j,k} + \Delta x_{i-1,j,k}},
  \end{eqnarray}
  where $\Delta x_{i-1,j,k} $ and $\Delta x_{i+1,j,k}$ denote the distance in the $x$-direction from the center of the cell to the center of the left and right  neighbor cells, respectively. They take into account the level of the neighbor cell.

  If the neighbor cell is coarser, we interpolate the pressure at the fine level using a slope limiter to compute the gradient. This method is similar to what is used in the hydrodynamical solver of {\ttfamily RAMSES}, with a loss of one order of accuracy at level interface. In certain conditions this method maintains a second-order accuracy of the scheme in presence of AMR (see Sect. \ref{sec:dustywave}).
  The case where the neighbor cell is at a finer level is never considered since fine-to-coarse fluxes are already computed during the update of the finer level.
Finally, the expression of the $x$-component of $\vec{w}^n_{i,j,k}$ is given by 
  \begin{eqnarray}
  {w_{x}}^n_{i,j,k} = \frac {{T_{\mathrm{s}}}_{i,j,k}^n\grad_{x} {P_{\mathrm{g}}}_{i,j,k}^n }{\rho_{i,j,k}^n}.
    \end{eqnarray}

  \paragraph{Predictor step:}

 During the predictor step, the dust density is estimated at the left and right interfaces with a simple finite difference method. It uses slope limiters to preserve the total variation diminishing property of the scheme \citep[TVD,][]{1983JCoPh..49..357H}. 
The centered value of the dust density is estimated at half timesteps as
\begin{eqnarray}
{\rho_{\mathrm{d}}}_{i,j,k}^{n + 1/2} &=& {\rho_{\mathrm{d}}}_{i,j,k}^n \nonumber \\ & -& \frac{\Delta t}{2 \Delta x} \sum_{{\sigma}=x,y,z} \left( {w_{\sigma}}^n_{i,j,k} \Delta_{\sigma} {\rho_{\mathrm{d}}}_{i,j,k}^n +{\rho_{\mathrm{d}}}_{i,j,k}^n\Delta_{\sigma}  {w_{\sigma}}^n_{i,j,k} \right)  ,
\end{eqnarray}
where  $\Delta_\sigma {\rho_{\mathrm{d}}}_{i,j,k}^n$ and $ \Delta_{\sigma}  {w_{\sigma}}^n_{i,j,k}$ are the TVD  variations of   $\rho_{\mathrm{d}}$ and $w_{\sigma}$ in the direction $\sigma$.  After the prediction of ${\rho_{\mathrm{d}}}_{i,j,k}^{n + 1/2}$, we interpolate   $\rho_{\mathrm{d}}$ and $w_{\sigma}$  at the interfaces. 

Let us consider the interface in the $x$-direction. The left and right interface values, denoted by the $L$ and $R$ subscripts, are given by

\begin{eqnarray}
{\rho_{L}}_{i,j,k}&=&  {\rho_{\mathrm{d}}}_{i,j,k}^{n + 1/2} - \frac{\Delta_{x} {\rho_{\mathrm{d}}}^n_{i,j,k}}{2}. \nonumber \\
{\rho_{R}}_{i,j,k}&=&  {\rho_{\mathrm{d}}}_{i,j,k}^{n + 1/2} + \frac{\Delta_{x} {\rho_{\mathrm{d}}}^n_{i,j,k}}{2 }, \nonumber \\
{{w_{x}}_{L}}_{i,j,k}&=&  {w_{x}}_{i,j,k}^{n} - \frac{\Delta_{x} {w_{x}}^n_{i,j,k}}{2 }, \nonumber \\
{{w_{x}}_{R}}_{i,j,k}&=&  {w_{x}}_{i,j,k}^{n} + \frac{\Delta_{x} {w_{x}}^n_{i,j,k}}{2}.
\end{eqnarray}

The bottom, top, back, and front  states are estimated in a similar way.  We do not perform a predictor step in time for the differential advection velocity because the current scheme  is sufficient to get the second-order convergence (see Sect. \ref{sec:dustywave}).
  \paragraph{Corrective step:}
  The correction operation  consists in computing the fluxes at the interface using the left and right predicted values, respectively.  We consider the left interface of the cell $i,j,k$ in the $x$-direction.  To avoid issues with the velocity discontinuities at the interfaces \citep[e.g., due to spurious pressure jumps,][]{2009arXiv0909.5426S}, we impose a unique advection velocity 
  \begin{eqnarray}
  w_{i-1/2,j,k}^{n + 1/2} = \frac{{w_L}_{i,j,k}^{n + 1/2}+{w_R}_{i-1,j,k}^{n + 1/2}}{2}.
  \end{eqnarray}
  This average does not appear to be critical for the second-order accuracy of the scheme but leads to a better convergence and is consistent with the upwind implementation in {\ttfamily RAMSES}.
 Our scheme uses the upwind method \citep{UPWIND} to estimate the flux, sufficient to get the second-order accuracy in space. It writes as 
\begin{eqnarray}
 \mathcal{F}^{n+1/2}_{i-1/2,j,k}  &=& \mathrm{max}\left(w_{i-1/2,j,k}^{n + 1/2} {\rho_{\mathrm{d},L}}_{i,j,k}^{n + 1/2},0\right) \nonumber \\ &+& \mathrm{min}\left(w_{i-1/2,j,k}^{n + 1/2} {\rho_{\mathrm{d},R}}_{i-1,j,k}^{n + 1/2},0\right).
\end{eqnarray}
   We note that several other approximate Riemann solvers can be used in our implementation as well, such as Lax-Wendroff \citep{doi:10.1002/cpa.3160130205} or Harten-Lax-van Leer \citep[][hereafter HLL]{articleHLL}.
The fluxes in the other directions $\mathcal{F}^{n+1/2}_{i,j-1/2,k}$ and  $\mathcal{F}^{n+1/2}_{i,j,k-1/2}$ are estimated in a similar way. The dust density is finally updated according to
  
\begin{eqnarray}
\label{eq:goddust}
{\rho_{\mathrm{d}}}_{i,j,k}^{n + 1} = {\rho_{\mathrm{d}}}_{i,j,k}^{n}& -& \frac{\Delta t}{\Delta x} \left(\mathcal {F}^{n+1/2}_{i+ 1/2,j,k} -\mathcal {F}^{n+1/2}_{i- 1/2,j,k}\right) \nonumber\\ &-&\frac{\Delta t}{\Delta x}  \left(\mathcal {F}^{n+1/2}_{i,j+ 1/2,k} -\mathcal {F}^{n+1/2}_{i,j- 1/2,k}\right) \nonumber\\ &-&\frac{\Delta t}{\Delta x} \left(\mathcal {F}^{n+1/2}_{i,j,k+ 1/2} -\mathcal {F}^{n+1/2}_{i,j,k- 1/2}\right).
\end{eqnarray}
 
\paragraph{Energy :}
 Similarly  and simultaneously with the dust diffusion step, the energy is updated after the hydrodynamical step to account for the energy component of $\mathbb{F}_{\mathrm{\Delta}}$. It is computed using the same scheme as the dust density with the velocity computed with Eq.\ref{eq:wbackk} and the internal energy instead of the dust density. The fluxes are then added to the state vector similarly to Eq. \ref{eq:goddust} with the total energy instead of the dust density.

   \subsection{Time-stepping}
   \label{sec:stab}
   
   Since the dust diffusion step consists in solving an advection equation explicitly, the scheme stability is achieved if the timestep $\Delta t$ verifies the split CFL condition
   \begin{eqnarray}
   \Delta t < \Delta t _{\mathrm{dust}} \equiv C_{\mathrm{CFL}}  \frac{\Delta x}{\sum_{\sigma=x,y,z}\left|\mathrm{max}(w_{\sigma},w_{\mathrm{g,b}\sigma})\right|},
   \end{eqnarray}
   where $C_{\mathrm{CFL}}<1$ is a safety factor.  In addition, the hydrodynamical step imposes another stability condition that writes
    \begin{eqnarray}
    \Delta t < C_{\mathrm{CFL}} \frac{\Delta x}{|c_{\mathrm{s}}|+\sum_{\sigma=x,y,z}\left|v_{\sigma}\right|},
   \end{eqnarray}
   where $v_{\sigma}$ is the mixture velocity in the direction $\sigma$. Dust diffusion is stable without intervention on the timestep as long as
      \begin{eqnarray}
      {|c_{\mathrm{s}}|+\sum_{\sigma=x,y,z}\left|v_{\sigma}\right|} >{\sum_{\sigma=x,y,z}\left|\mathrm{max}(w_{\sigma},w_{\mathrm{g,b}\sigma})\right|} .
   \end{eqnarray}
   If the former condition is not verified, which is the case when the pressure gradient is steep, dust diffusion constrains the timestep. In this case, we impose the dust timestep instead of the hydrodynamical  one. 

\section{Validation tests}

To benchmark the implementation of dust dynamics in {\ttfamily RAMSES}, we run the canonical tests for gas and dust mixtures, \textsc{dustydiffuse}, \textsc{dustyshock}, \textsc{dustywave}  and the \textsc{settling} test. We also test this advection solver with the \textsc{dustyadvect} test. 

\label{sec:tests}
\subsection{Dustyadvect}

\begin{figure*}[h!]
       \centering
          \includegraphics[scale=0.8]{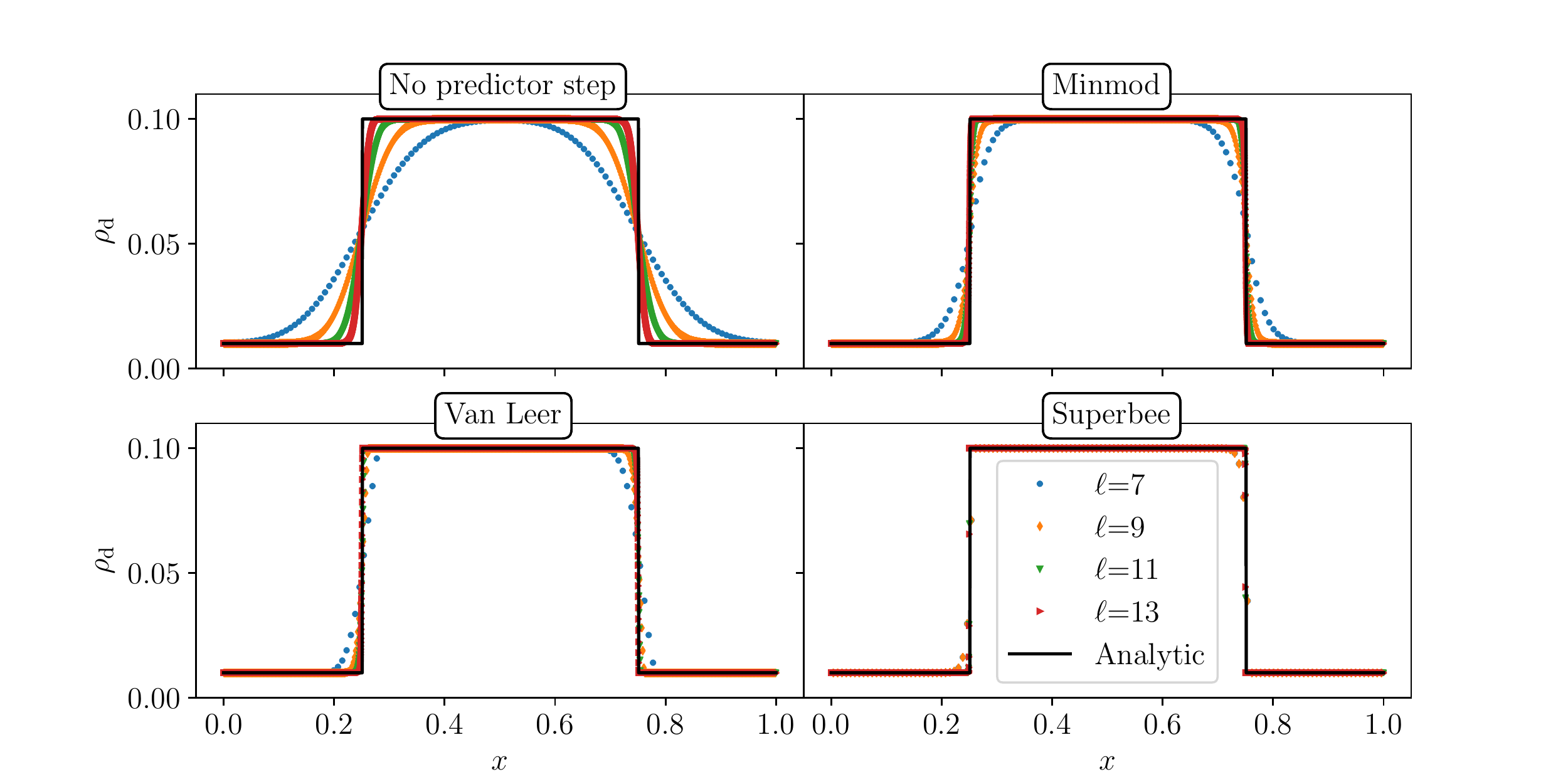}
      \caption{\textsc{dustyadvect} tests using the function  $f_1$ (Eq. (\ref{eq:f1})). Dust density after one on period various grids ($\ell=7$ in blue, $\ell=9$ in orange, $\ell=11$ in green, $\ell=13$ in red) as a function of the position compared with the analytic solution (black solid line). We present this test using four different slope limiters.  (Top left) No predictor step. (Top right) Minmod slope limiter. (Bottom left) Van-Leer slope limiter. (Bottom right) Superbee slope limiter. }
       \label{fig:adv1} 
\end{figure*}

\begin{figure}[h!]
       \centering
          \includegraphics[scale=0.7]{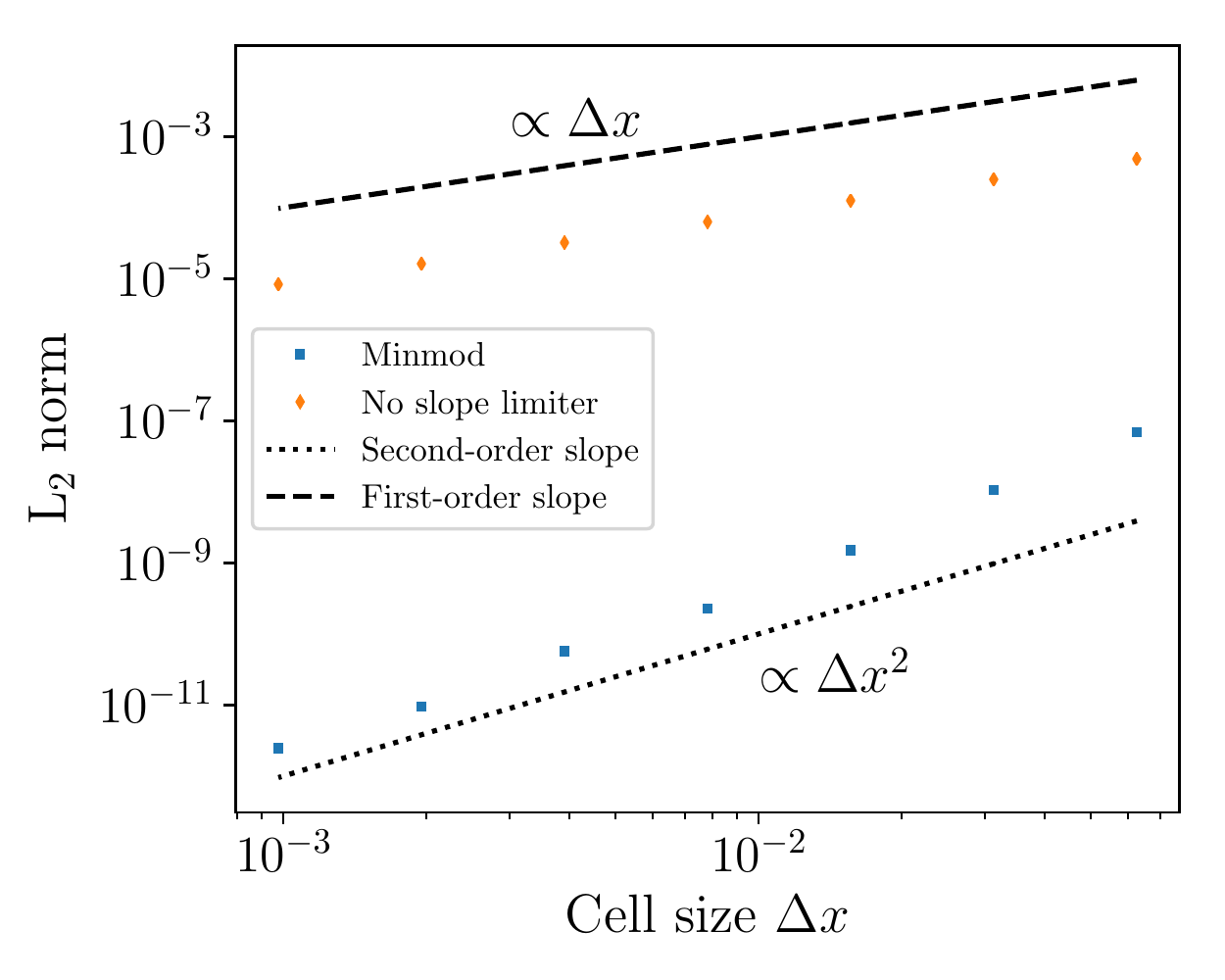}
      \caption{\textsc{dustyadvect} tests  with an initial condition given by the function  $f_2$ (Eq. (\ref{eq:f2})). $L_2$ (Eq. (\ref{eq:l2})) norm as a function of the cell size for the scheme using the Minmod slope limiter (blue squares) and without predictor step (orange diamonds). The results are compared with a first-order slope (dashed line) and a second-order slope (dotted  line). }
       \label{fig:adv2} 
\end{figure}
The scheme convergence and behavior at discontinuities is examined with 1D advection tests. In these \textsc{dustyadvect} tests, the advection velocity $w_x$ is constant and the hydrodynamical update deactivated. 
It consists in solving the following equation 
\begin{eqnarray}
\frac{\partial \rho_{\mathrm{d}}}{\partial t} = - w_x \frac{\partial \rho_{\mathrm{d}}}{\partial x}.
\label{eq:advcons}  
\end{eqnarray}
Considering an initial condition $\rho_{\mathrm{d}}(x,0)=f(x)$, $f(x)$ having a period of the size of the box $L=1$, the analytic solution at time $t$ is simply given by $\rho_{\mathrm{d}}(x,t)=f(x-w_x t)$. In the remainder of the section, $w_x=1$.  

At first, four \textsc{dustyadvect} tests are performed. We impose the initial function $f_1$, which writes $ \forall x \in \left[0,L\right]$
\begin{eqnarray}
f_1\left(x\right) = \left\{
    \begin{array}{ll}
        0.1 & \mbox{if } \frac{L}{4}<x<\frac{3L}{4}, \\
        0.01& \mbox{otherwise.}
    \end{array}
\right.
\label{eq:f1}
\end{eqnarray} 
  In the first test, no predictor step is operated. Different slope limiters are then used for the three other tests in the predictor step; Minmod \citep{1986AnRFM..18..337R}, Van-Leer  \citep{1974JCoPh..14..361V}, and Superbee  \citep{1986AnRFM..18..337R}. Uniform grids of resolutions ranging from $\ell=7$  (128 cells) to $\ell=13$ (8192 cells) are considered.  Extremely small timesteps compared with the CFL condition $\Delta t = 8 \times 10^{-6}$ are imposed to ensure that the spatial error dominates. 

 Figure \ref{fig:adv1} shows the outcome of these tests after one period, at $t=1$. As expected, the quality of the results strongly depends on the slope limiter. Without the predictor step (no slope limiter) the solver is simply a first-order centered upwind scheme. In this case a larger resolution is required to achieve the same accuracy as in the three other tests. As expected, the Van-Leer  and Superbee  slope limiters give more accurate results than the Minmod  test but at the cost of a lack of symmetry. The  Minmod  slope limiter was therefore chosen as a good compromise to achieve a satisfying precision and to preserve symmetry. 

Another series of \textsc{dustyadvect} tests are performed for different resolutions, using the Minmod slope limiter and without a predictor step. A smooth initial Gaussian-like function $f_2$ is imposed to quantify the truncation error in space of the scheme which writes,  $\forall x \in \left[0,L\right]$,
\begin{eqnarray}
f_2\left(x\right)& = &0.01 + 0.1 \exp\left[{-{\left(\frac{x-L/2}{L/4}\right)}^2}\right].
\label{eq:f2}
\end{eqnarray}

 We use  a very small timestep $\Delta t = 10^{-8}$ to minimize the truncation errors in time compared with spatial errors. The results are compared at the same time  $t = 0.01$ with the analytic solution using the $L_2$ norm
\begin{eqnarray}
L_{2}= \sqrt{\frac{\sum_{i=1}^{N_{\mathrm{cell}}}\left|{{\rho_{\mathrm{d}}}_{\mathrm{RESULTS},i}^n}-{{\rho_{\mathrm{d}}}_{\mathrm{ANALYTIC},i}^n}\right|^2}{N_{\mathrm{cell}}}}.
\label{eq:l2}
\end{eqnarray}

 Figure \ref{fig:adv2} shows the evolution of the $L_2$ norm with (blue squares) and without (orange diamonds) a prediction step as a function of the size of the cells for the Gaussian-like test.
As expected, without prediction, the scheme has only a first-order accuracy in space, while it is a second-order scheme when the full predictor-corrector scheme with slope limiters are used. 

\subsection{Dustydiffuse}
\label{subsec:dustdydiff}
\begin{figure}[h!]
       \centering
          \includegraphics[scale=0.72]{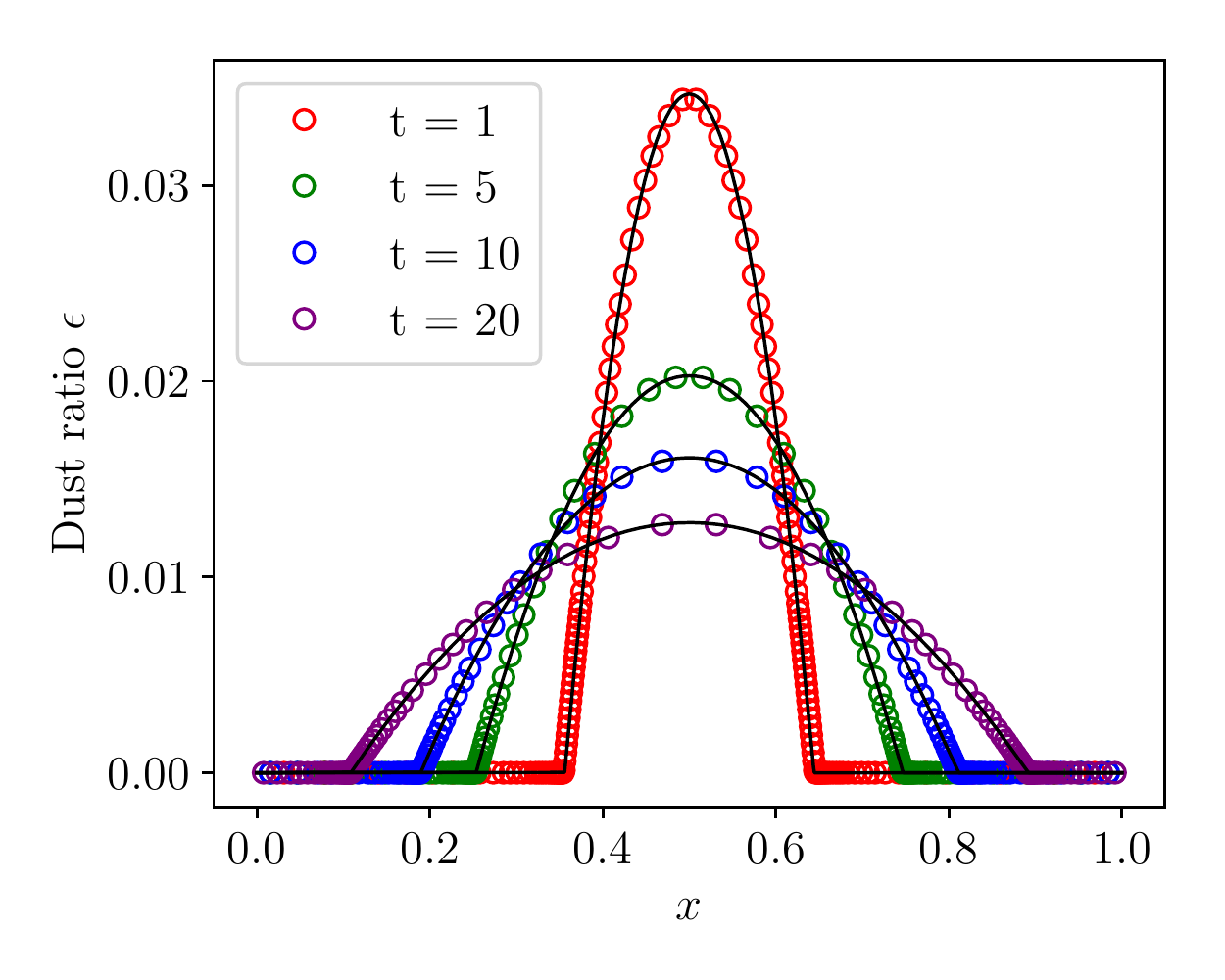}
      \caption{\textsc{Dustydiffuse} tests. Dust ratio as a function of the position at $t=1$ (red), $t=5$ (green), $t=10$ (blue), $t=20$ (purple) compared with the exact solution (black solid lines).}
       \label{fig:diffamr} 
\end{figure}

 When the hydrodynamical variables (pressure and dust ratio excluded) remain constant, Eq. (\ref{eq:eqdiff}) behaves as a non-linear diffusion equation. Pure dust diffusion tests can thus be performed. In the \textsc{dustydiffuse} tests \citep{2015MNRAS.454.2320P}, the hydrodynamical step is deactivated and  $t_{\mathrm{s}}$ and $c_{\mathrm{s}} $ are set constant as well. Therefore, we only solve the following equation:
\begin{eqnarray}
\frac{\partial \rho_{\mathrm{d}}}{\partial t} = \frac{\partial \rho_{\mathrm{d}}w_x}{\partial x}.
\label{eq:bar}  
\end{eqnarray}
An isothermal equation of state $ P_{\mathrm{g}} =c_{\mathrm{s}}^2(1- \epsilon) \rho$ is considered.  The advection velocity is then given by
\begin{eqnarray}
w_x=  -t_{\mathrm{s}}  c_{\mathrm{s}}^2 \frac {\partial \epsilon}{\partial x};
\label{eq:w_x_iso}  
\end{eqnarray}
    $\rho$ being  constant, the equation can be written as the a non-linear diffusion equation 
\begin{eqnarray}
\frac{\partial \epsilon}{\partial t} = t_{\mathrm{s}}  c_{\mathrm{s}}^2  \frac{\partial }{\partial x} \left(\epsilon\frac {\partial \epsilon}{\partial x}\right).
\label{eq:bar2}  
\end{eqnarray}
Equation (\ref{eq:bar2}) has a self similar solution known as the Barenblatt-Pattle solution \citep{Barenb} that writes
\begin{eqnarray}
\epsilon (t,x,C) = (t_{\mathrm{s}} c_{\mathrm{s}}^2 t)^{-1/3}\left(C- \frac{1}6\frac{x^2}{(t_{\mathrm{s}} c_{\mathrm{s}}^2 t)^{2/3}}\right),
\label{eq:solbar}  
\end{eqnarray}
where $C$ is a constant depending on the initial conditions. We consider the following initial profile 
\begin{eqnarray}
\epsilon (t_0,x) = \epsilon_0 \left( 1 - \left(\frac{x}{x_{\mathrm{c}}}\right)^2\right) ,  
\end{eqnarray}
which is consistent with  Eq. (\ref{eq:solbar}) if  
\begin{eqnarray*}
 t_0& =& \frac{C^3}{t_{\mathrm{s}}  c_{\mathrm{s}}^2 \epsilon_0^3},  \nonumber \\
 C&=& \left(\frac{\epsilon_0 x_{\mathrm{c}}}{\sqrt{6}}\right)^{2/3}.\end{eqnarray*}

We additionally set  $\rho=1$, $t_{\mathrm{s}}=0.1$, $\epsilon_0=0.1$ and  $c_{\mathrm{s}} = 1$ and an AMR grid of $\ell_{\mathrm{min}}= 4$ and with $\ell_{\mathrm{max} } =10$. The refinement criterion, based on the dust density gradient, forbids a variation of more than $5 \%$ between two cells.  

 Figure \ref{fig:diffamr} shows a comparison between the outcome of the tests and the analytic solutions at $t=1$, $t=5$, $t=10$ and $t=20$. At each time the numerical results agree with the exact solution to a precision of less then $1\%$ in $L_2$ norm. Even though our scheme is fundamentally designed for advection, it is also efficient at handling diffusion problems.

\subsection{Dustyshock}

 \begin{figure*}[h!]
       \centering
          \includegraphics[scale=0.6]{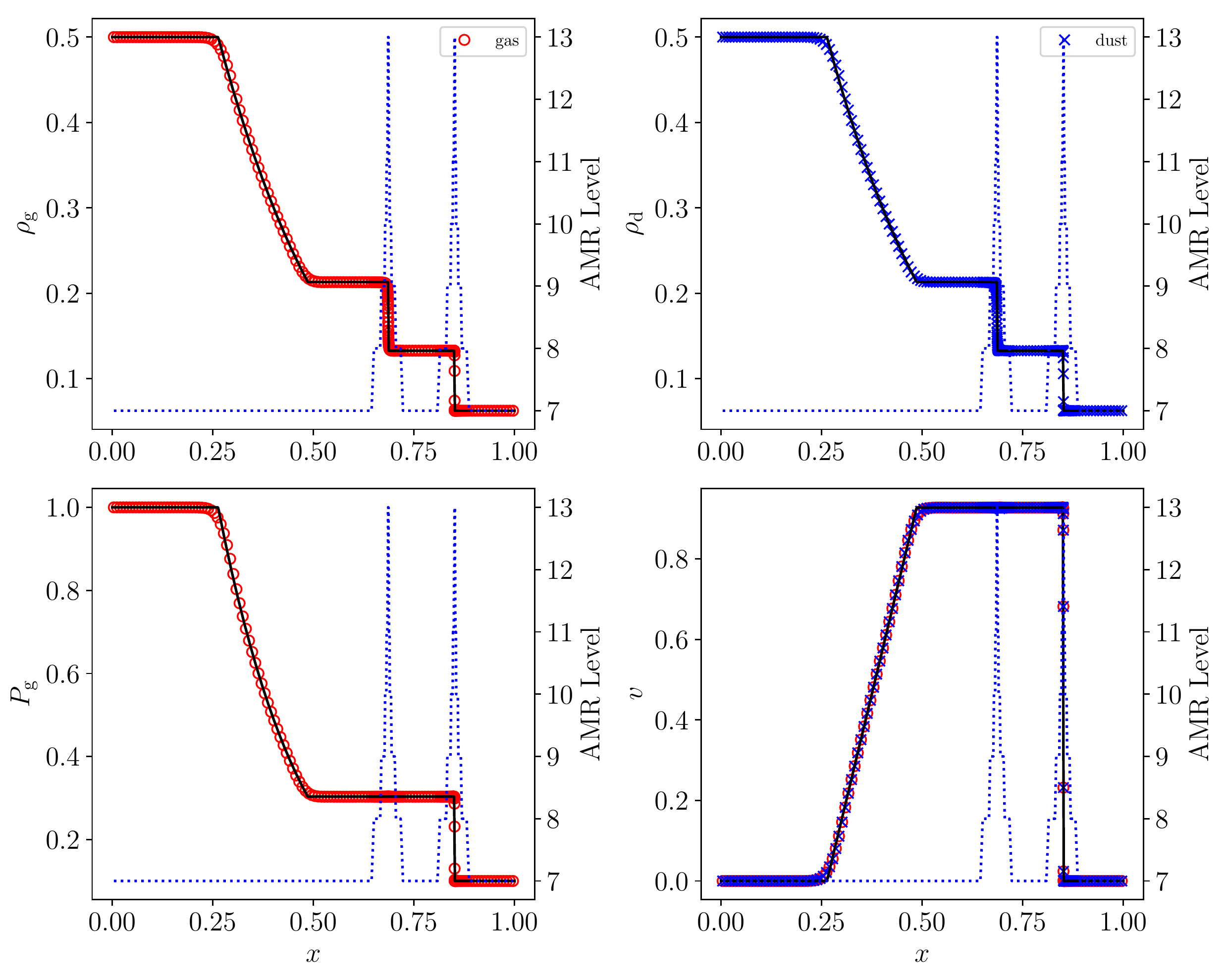}
      \caption{\textsc{Dustyshock} with $\epsilon={0.5}$. (Top left) Gas density as a function of position. (Top right) Same but for dust. (Bottom left) Gas pressure. (Bottom right) Gas and dust velocities.  The AMR level (right axis) is represented with dotted blue lines. The analytic solution is given by the black solid lines. Red circles and blue crosses indicate gas and dust numerical quantities, respectively.}
      \label{fig:dustyshock}
      \end{figure*}

Another canonical test for gas and dust mixtures consists of 1D hydrodynamical shocks.  For strongly coupled mixtures,  these so-called \textsc{dustyshock} tests are closely approximated by the same analytic solution as the usual Sod test \citep{1978JCoPh..27....1S}, but with the modified sound speed $\tilde{c}_{\mathrm{s}}$
\begin{eqnarray*}
\tilde{c}_{\mathrm{s}} ={c_{\mathrm{s}}} \sqrt{1- \epsilon_0},
\label{eq:dustycs}
\end{eqnarray*}  
$\epsilon_0$ being the initial dust ratio. The dust ratio $\epsilon$ remains almost constant through the shock; however, pressure bumps must occur where there is either a pressure or density gradient. 

The same prescription as \cite{2014MNRAS.440.2136L} for the stopping time is used. It writes 
 \begin{eqnarray*}
   t_{\mathrm{s}} = \frac{\epsilon (1- \epsilon) \rho}{K},
   \end{eqnarray*}
where $K$ is  the drag coefficient defined previously. This prescription is consistent with the expression of the stopping time  presented before is convenient for tests since it allows us simply parametrize the coupling between the gas and dust phase. 

A  \textsc{dustyshock} test is performed on an AMR grid with $\ell_{\mathrm{min}}=4$ and  $\ell_{\mathrm{max}}=13$ with a high initial dust ratio. The grid is refined with a criterion that forbids dust density variations of more than $5\%$ between two neighbor cells. Two distinct regions, representing the left and right half of the box, are set with different initial conditions given by
\begin{eqnarray*}
 \left(\rho_0, \vec{v}_0, {P_{\mathrm{g}}}_0, \epsilon_0 \right)_{\mathrm{left}}&=& \left(1,0,1,{0.5}\right)  \nonumber \\
 \left(\rho_0, \vec{v}_0, {P_{\mathrm{g}}}_0, \epsilon_0 \right)_{\mathrm{right}}&=& \left(0.125,0,0.1,{0.5}\right).
 \end{eqnarray*}
  Finally, a drag coefficient $K=1000$  and an adiabatic index $\gamma =1.4$ are imposed. 

Figure \ref{fig:dustyshock} shows the gas and dust densities, the velocity, and the gas pressure as a function of the position at $t=0.2$. The Sod solution with the modified sound speed is very well recovered.

\subsection{Dustywave}
\label{sec:dustywave}
\begin{figure*}[h!]
       \centering
          \includegraphics[scale=0.7]{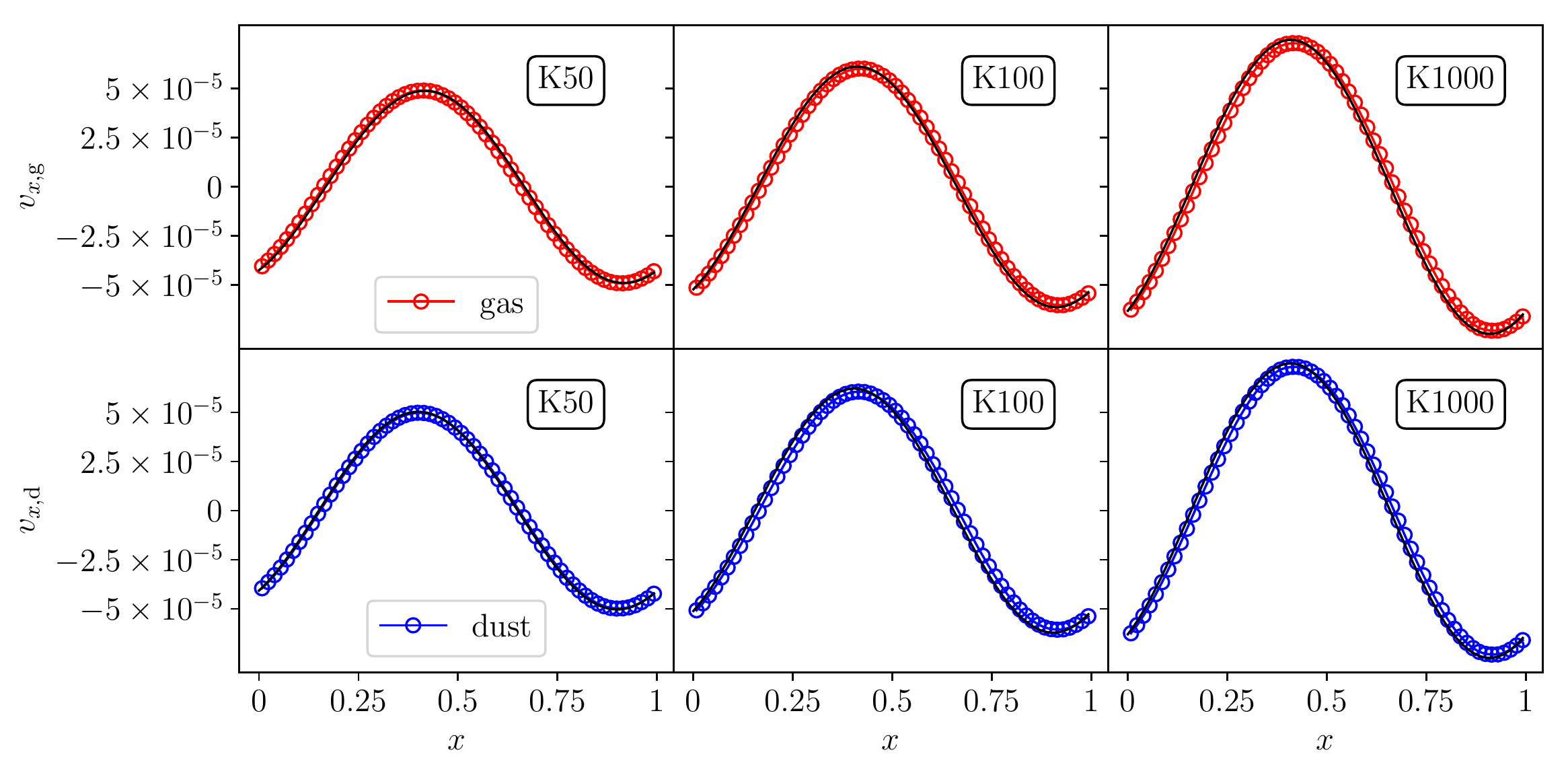}
      \caption{\textsc{dustywave} test. Velocity of the gas (top) and dust (bottom) phase  as a function of the position at $t=4.5$ for the  K50, K100 and K1000 tests (from left to right) compared with the analytical solution (black lines)  given by  \cite{2011MNRAS.418.1491L}.}
      \label{fig:dustywavev}
      \end{figure*}
 \begin{figure*}[h!]
       \centering
          \includegraphics[scale=0.7]{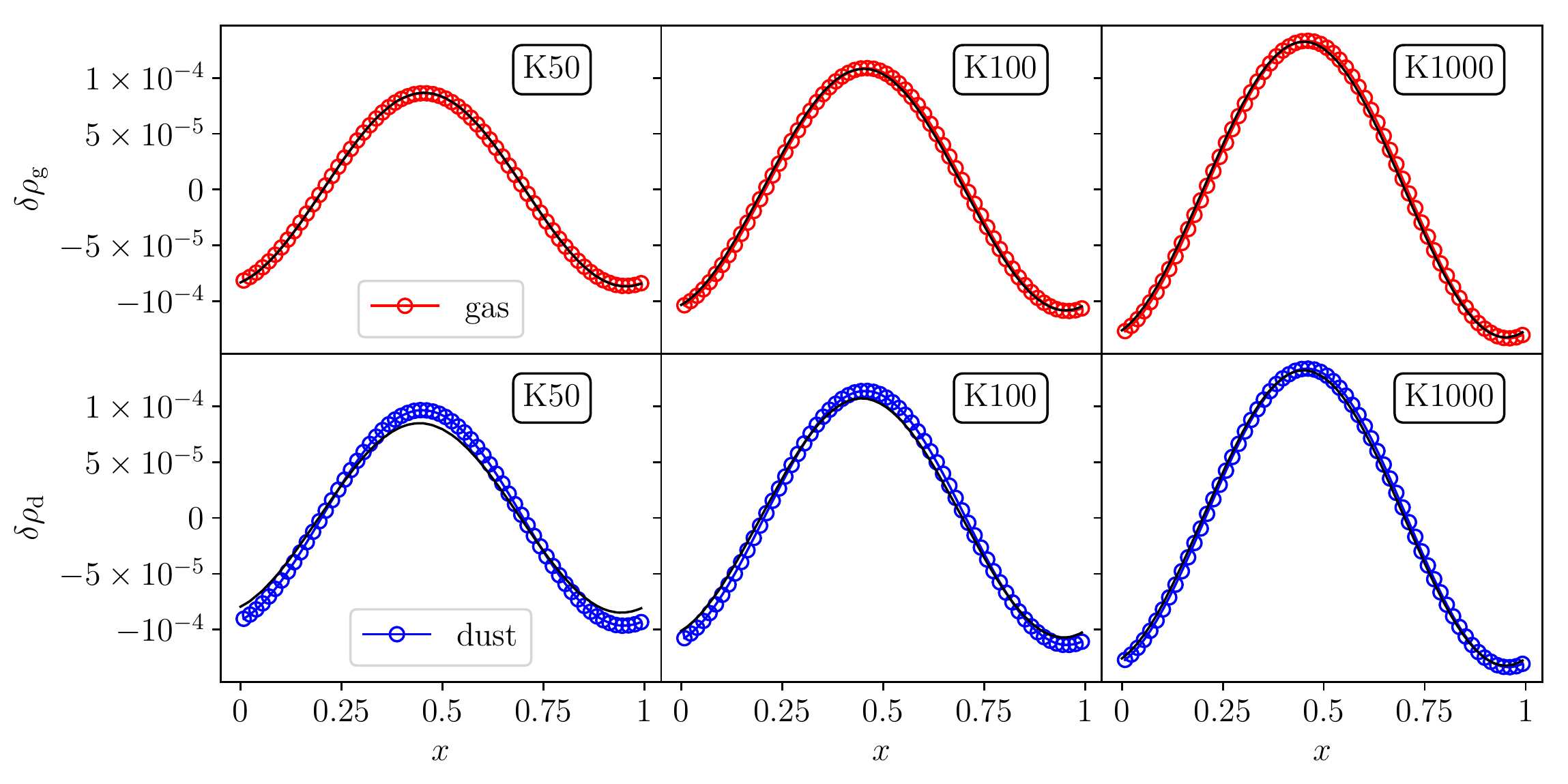}
      \caption{\textsc{dustywave} test. Same as in Fig. \ref{fig:dustywavev} but for the density perturbations.}
      \label{fig:dustywaved}
      \end{figure*}
\begin{figure}[t]
       \centering
          \includegraphics[scale=0.7]{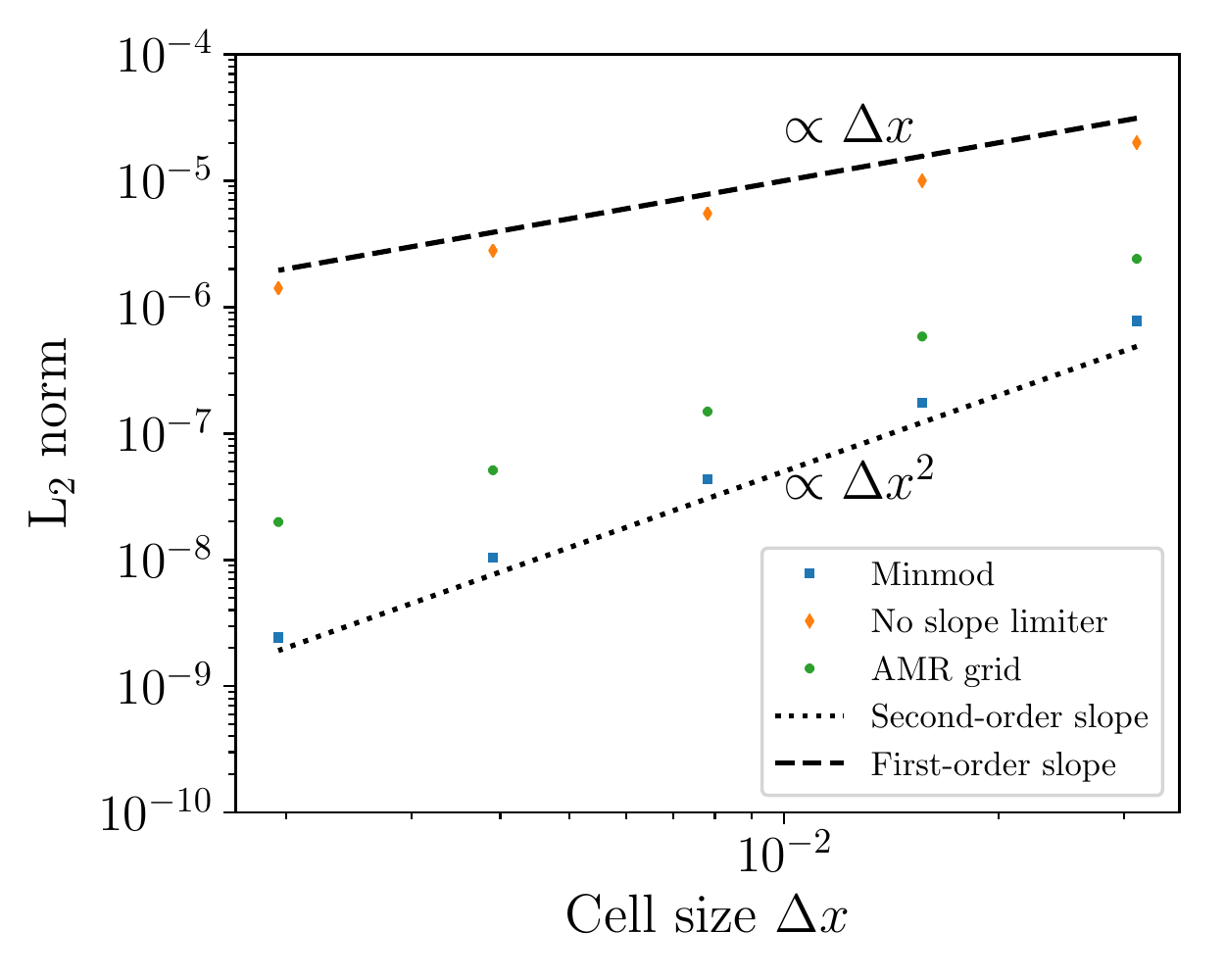}
      \caption{\textsc{dustywave} numerical convergence tests. $L_2$ norm as a function of the  minimum  cell size for the scheme using the Minmod slope limiter (blue squares), without predictor step (orange diamonds)  and with AMR (green circles) . The results are compared with a first-order slope (dashed line) and a second-order slope (dotted line). }
       \label{fig:dwaveconv} 
\end{figure}      
      
The 1D \textsc{dustywave} test \citep{2011MNRAS.418.1491L} consists in following the evolution of an isothermal sound wave in a gas and dust mixture. Small periodic perturbations on the Eqs. (\ref{eq:fullmono}) on the density, the dust ratio, the pressure, and the velocity are imposed
\begin{eqnarray*}
\rho &=&\rho_0+ \delta \rho_0, \nonumber \\
\epsilon &=& \epsilon_0 + \delta \epsilon_0, \nonumber \\
{P_{\mathrm{g}}}&=& {P_{\mathrm{g}}}_0 + \delta {P_{\mathrm{g}}}_0, \nonumber \\
{v_x} &=&0 +\delta {v_x}_0,
\end{eqnarray*}
 where the index $0$ and  the symbol $\delta$  indicate respectively  the initial uniform quantities and the perturbations.  \cite{2011MNRAS.418.1491L} provide the analytic solution for this test. They find that the sound waves propagate with the modified sound speed $\tilde{c}_{\mathrm{s}}$  and are damped because of the dust inertia. Large grains damp these sound waves faster than small grains as they are more massive.
 
 Three \textsc{dustywave} tests are performed using different drag coefficients $K=50$ (K50,  strong back-reaction regime), $K=100$ (K100) and $K=1000$ (K1000,weak back-reaction regime).  The initial perturbations are in the form
 \begin{eqnarray*}
  \delta\rho_0 &=& \rho_0 \delta  \sin \left(2 \pi \frac{x}{L}\right), \nonumber \\
\delta {v_x}_0 &= &v_0 \delta \sin \left(2 \pi \frac{x}{L}\right),\nonumber \\
 \delta {P_{\mathrm{g}}}_0 &=& \left(1-\epsilon_0\right) c_{\mathrm{s},0}^2 \delta \rho_0,\nonumber \\
\delta \epsilon_0 &=& 0,
\end{eqnarray*}
 where $x$ is the position in the box, $L$ is the box length, $ c_{\mathrm{s},0}$ is the initial sound speed, and $\delta$ is a parameter that sets the amplitude of the perturbation.
 The initial uniform state is set such as that
 \begin{eqnarray*}
 \rho_0 &=& 2, \nonumber \\
 \epsilon_0 &=& 0.5,\nonumber \\
 v_0&=&1, \nonumber  \\
 c_{\mathrm{s},0} &=& 1.
  \end{eqnarray*}     
 The adiabatic index of the gas is $\gamma = 1.000001 $\footnote{ Taking exactly $\gamma=1$ is not possible with {\ttfamily RAMSES} as the internal energy is computed as $\frac{P_{\mathrm{g}}}{\gamma -1}$} to simulate an isothermal soundwave propagation.  The initial perturbation has a relative amplitude  $\delta =10^{-4}$. The simulation box has a size $L=1.0$ and the grid is taken as uniform with $64$ cells. The timestep $\Delta t =10^{-4}$ is the same for the three tests  and respects the stability condition for the considered drag coefficients. 
  
  Figures \ref{fig:dustywavev} and \ref{fig:dustywaved} show the velocity $v_x$ and the perturbation density $\delta \rho=\rho-\rho_0$ for both gas and dust at $t=4.5$ for these three tests.  The amplitude of the damping increases with a decreasing $K$  and the results are increasingly less accurate as the errors due to the diffusion approximation increase, consistently with the theory \citep{2011MNRAS.418.1491L,2015MNRAS.454.2320P}. Larger grains, i.e, with smaller $K$, have more inertia and allow an efficient damping of the gas. 
  
   We perform \textsc{dustywave} tests on uniform grids with resolutions ranging from $\ell=5$ to $\ell=9$, with a small timestep $\Delta t =10^{-5}$  and for $K=50$.  We also perform runs on AMR grids with coarse resolutions ranging from $\ell_{\mathrm{min}}=4$ to $\ell_{\mathrm{min}}=8$. For these tests, the cells for $x \in \left[0.25,0.75 \right]$ have a  level of refinement $\ell_{\mathrm{max}}=\ell_{\mathrm{min+1}}$. We test the order of our scheme and measure the accuracy of numerical solution against a solution of reference obtained at very high resolution ($\ell=11$) in both time and space ($\Delta t =10^{-6}$). We do not compare the results with the analytic solution presented above as it not exact in the diffusion approximation. Figure  \ref{fig:dwaveconv} shows the  $L_2$ errors obtained when increasing the number of cells. The scheme is first-order in space without correction and second-order when using the Minmod limiter.  In the presence of AMR, the scheme keeps a second-order accuracy in space for low resolution but deteriorates at high resolutions. This is due to the estimate of the pressure gradient that is first-order at a refined interface. At high resolution, the error is smaller (almost two orders of magnitude) than the tests without a predictor step.
\subsection{ Disk settling }
\label{sec:sett}

\begin{figure}[t!]
       \centering
          \includegraphics[scale=0.7]{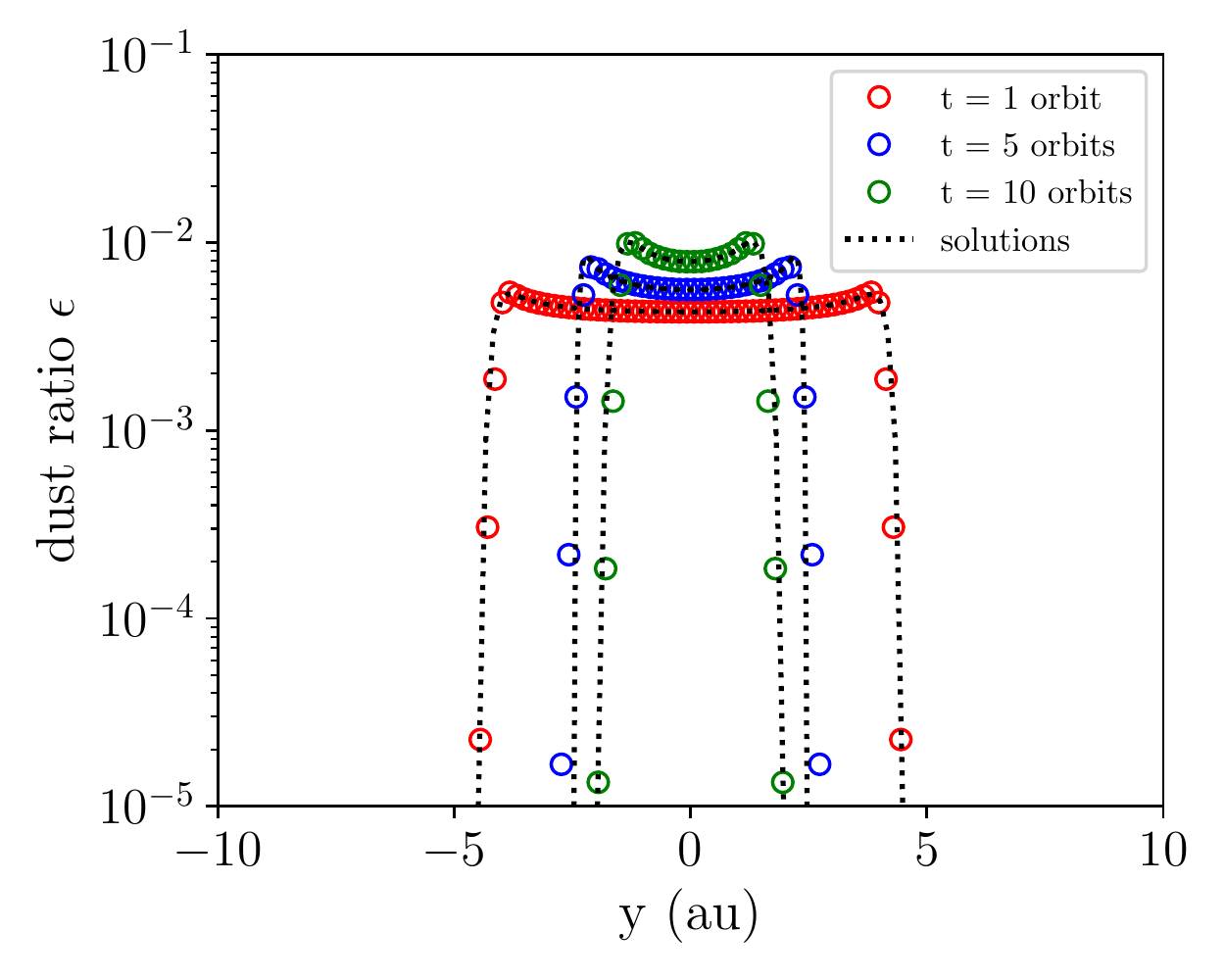}
      \caption{\textsc{settling} test on a uniform grid at level $\ell = 8$ with  millimeter-size dust grains. Dust ratio at t=1 (red circles), 5 (blue circles), and 10 orbits (green circles) as a function of the altitude.  }
       \label{fig:setunif} 
\end{figure}
 The \textsc{settling} test, introduced by \cite{2015MNRAS.454.2320P}, is designed to test the algorithm in realistic astrophysical conditions. It simulates the local settling of dust grains in a disk in hydrostatic equilibrium. As in \cite{2015MNRAS.454.2320P}, we set an analytic gravitational force \begin{eqnarray}
f_{\mathrm{grav}}= - \frac{ \mathcal{G} M_{\star} y }{{(y^2+r^2)}^{3/2}},\nonumber
\end{eqnarray} where $\mathcal{G}$ is the gravitational constant, $M_{\star}$ is the mass of the central star, $y$ is the altitude and $r$ the cylindrical radius at which the disk is simulated. The gas density at hydrostatic equilibrium for $ |y|\ll r $ is 
\begin{eqnarray}
\rho_{\mathrm{g}} = \rho_{\mathrm{g},0} e^{-\frac{y^2}{2 \mathcal{H}^2}}, \nonumber
\end{eqnarray}
where $\mathcal{H}$ is the local scale height of the disk, which is determined by the ratio $\mathcal{H}/r=0.05$. We then assume an isothermal equation of state where the   imposed soundspeed $c_{\mathrm{s}}$ is 
\begin{eqnarray}
c_{\mathrm{s}} = \mathcal{H} \Omega_{\mathrm{k}},
\end{eqnarray}
$\Omega_{\mathrm{k}}$ being the Keplerian angular velocity at the radius $r$.
Finally, a uniform initial dust ratio $\epsilon_0$ is imposed.

In the terminal velocity approximation, the analytic solution for the dust velocity is directly set by the pressure gradient that compensates the gravitational force, hence
\begin{eqnarray}
v_{\mathrm{d},y}= w_Y = - \mathrm{St}  \Omega_{\mathrm{k}} \frac{y}{(1+(y/r)^2)^{3/2}},
\end{eqnarray}
which is the limit at low Stokes number ($\mathrm{St} = \Omega_{\mathrm{k}} t_{\mathrm{s}}$ in this case) of the expression given by \cite{2018MNRAS.476.2186H}.
As the gas approximately remains in equilibrium, solving the settling problem consists in solving
\begin{eqnarray}
 \frac{\partial   \rho_{\mathrm{d}} }{\partial t}+\frac{\partial \rho_{\mathrm{d}} w_y }{\partial y}  = 0.
\end{eqnarray}
Even though the density can, in principle, be determined using the same method as the velocity, it relies on the hypothesis of an infinite dust reservoir that is inconsistent with our choice of boundaries. We choose to compare our results with a numerical solution as in \cite{2018MNRAS.476.2186H}. We use a similar Crank-Nicholson scheme to get this solution, except that it only solves the dust density equation using the analytic dust velocity.

We perform a 2D \textsc{settling} test at a disk orbiting around a solar mass star at radius of $50$~au. The simulation box has periodic boundary conditions and $L=20$~au, which is approximately $8\mathcal{H}$. As in \cite{2015MNRAS.454.2320P} and  \cite{2018MNRAS.476.2186H}, the initial mid-plane gas density $\rho_{\mathrm{g}} \approx 6 \times 10^{-13}$g cm$^{-3}$. We also set an initial dust ratio of $4.99 \times10^{-3}$ of millimeter grains with an intrinsic density of $\rho_{\mathrm{grain}}=3$~g cm$^{-3}$. The adiabatic index of the gas is $\gamma=5/3$.

A first test with a uniform grid at level $\ell=8$ is performed. Figure  \ref{fig:setunif} shows the dust ratio as a function of $y$. As can be seen, the results are essentially similar to those obtained in \cite{2015MNRAS.454.2320P} and  \cite{2018MNRAS.476.2186H}.
 \subsection{Multigrain}
 \subsubsection{Example 1 : Dustydiffuse}
 \begin{figure}[t]
       \centering
          \includegraphics[scale=0.7]{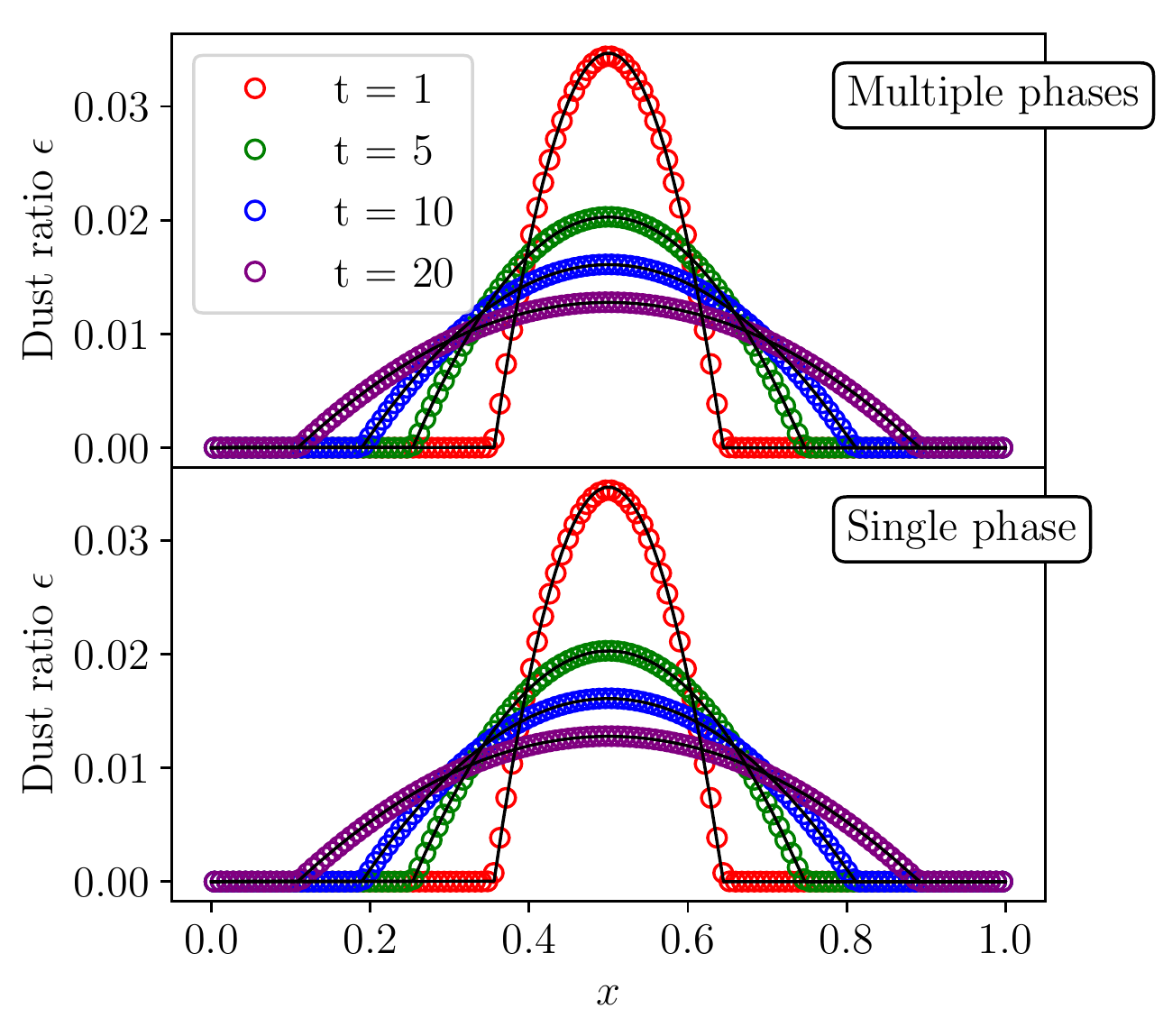}
      \caption{Comparison between  multiple phase with $\mathcal{N}=5$ (top) and an equivalent single phase (bottom) \textsc{dustydiffuse} test. Dust ratio as a function of the position at $t=1$ (red), $t=5$ (green), $t=10$ (blue) and $t=20$~s (purple) compared with the exact solution (black solid lines).}
       \label{fig:diffmulti} 
\end{figure}

 \begin{figure}[t]
       \centering
          \includegraphics[scale=0.7]{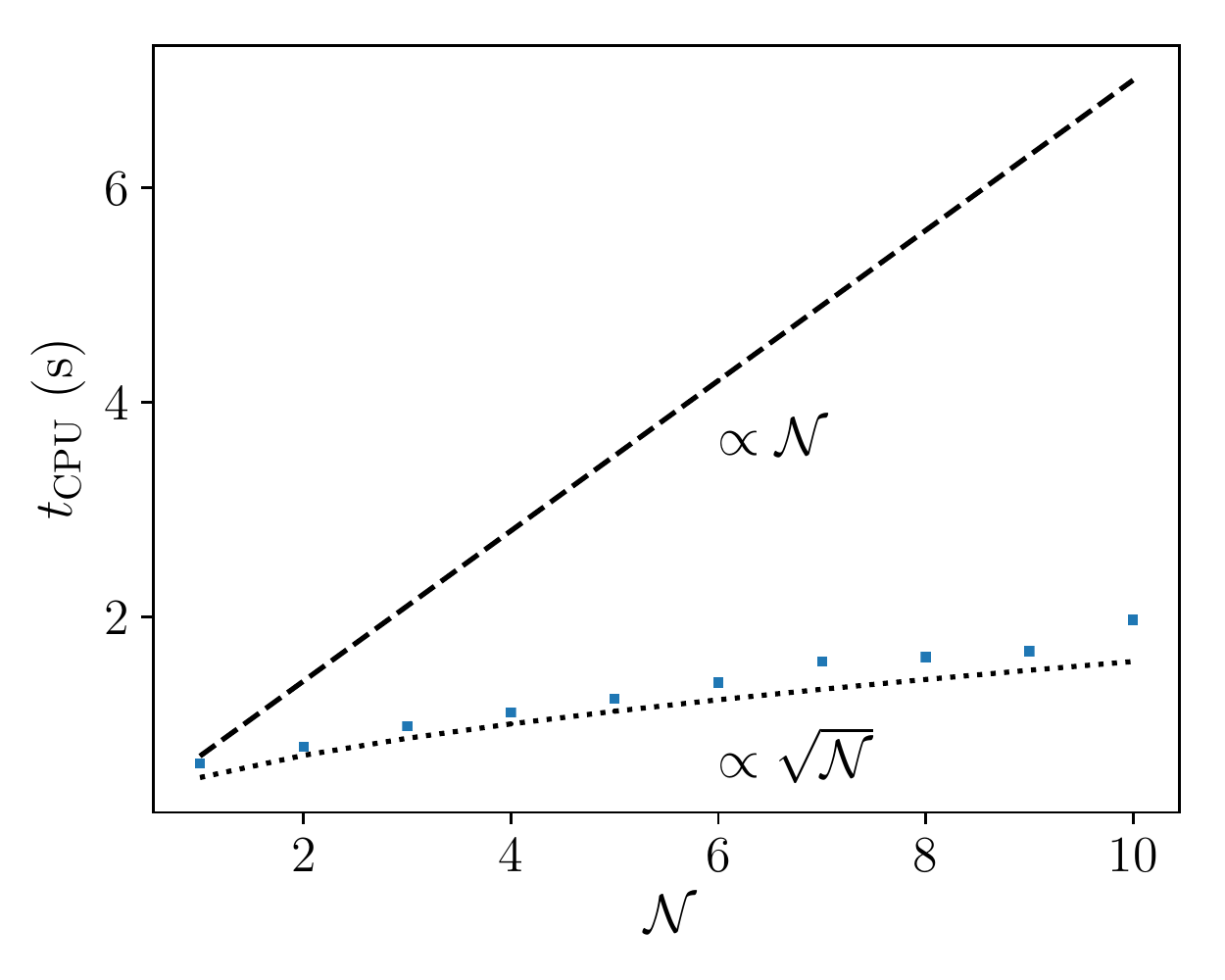}
      \caption{CPU time  $t_{\mathrm{CPU}}$ of the ten equivalent \textsc{dustydiffuse} tests as a function of the number of species $\mathcal{N}$.}
       \label{fig:TCPU} 
\end{figure}
 To benchmark the implementation of the multiple dust species in the code, ten \textsc{dustydiffuse} tests are performed on uniform grids of $\ell = 7 $. They are operated with one dust phase and with this same phase split with $\mathcal{N}\in \left[2,10\right]$ (\textsc{multigrain}).  All the dust bins have the same intrinsic properties. As all the tests are equivalent, it is expected that they give the same results. The parameters used are the same as in Sect. \ref{subsec:dustdydiff}. 
   
     Figure \ref{fig:diffmulti} shows the total dust density as a function of position with $\mathcal{N}=5$ (top) and $\mathcal{N}=1$ (bottom) at $t=1$, $t=5$, $t=10$, and $t=20$.  The results agree for the \textsc{multigrain} simulation and for the single phase simulation to machine precision.
     
     Figure \ref{fig:TCPU} shows the CPU time as a function of the number of species $\mathcal{N}$. We see that the \textsc{multigrain} simulations are not very expensive. The CPU time agrees more with a square root scaling with $\mathcal{N}$ than a linear one.  As discussed by \cite{2018MNRAS.476.2186H} with the \textsc{multigrain} algorithm in the {\ttfamily PHANTOM} code \citep{2017ascl.soft09002P}, the monofluid formalism is a highly computationally effective tool to treat multiple phases.
     
       \begin{figure}[h!]
       \centering
          \includegraphics[scale=0.5]{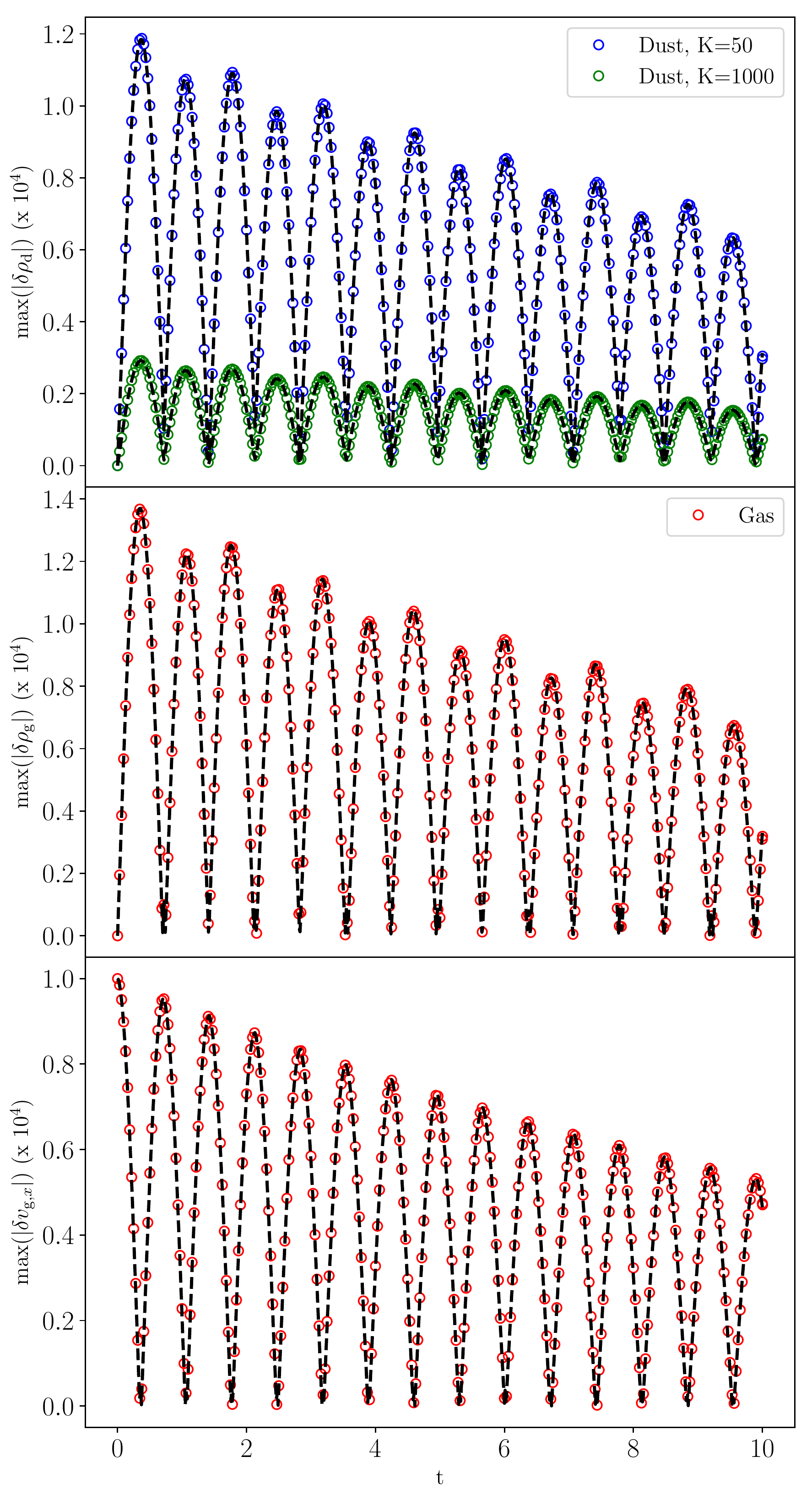}
      \caption{  \textsc{multigrain} \textsc{dustywave} test. The three panels show the evolution of the maximum amplitude of the perturbation for the dust densities (top), the gas density (middle), and the gas velocity (bottom). The $K=50$ and $K=1000$ dust phases are shown with blue and green circles respectively. The red circles represent the gas. The semi-analytic solution is given by the dashed black line.  }
       \label{fig:dustyamp} 
\end{figure} 
  \subsubsection{Example 2 : Dustywave}

   In this section, we test the cumulative back-reaction of dust on the gas and the interaction between dust species. To benchmark our \textsc{multigrain} implementation in a strong back-reaction regime we run a \textsc{dustywave} simulation with two dust species.  The initial barycentric velocity is a sinusoidal perturbation of amplitude $10^{-4}$, the other variables are not initially perturbed. We numerically integrate the linearly perturbed equations of gas and dust mixtures assuming solutions of the type $A(t) \exp(i k x)$  as in \cite{2014MNRAS.444.1940L} to obtain our  reference solution. The test is performed with a uniform grid of $\ell = 9$ and timesteps given by the CFL condition. In this test, we consider two dust phases and a total initial dust ratio of $0.5$. The first bin has a drag coefficient $K=50$ and an initial dust ratio of $0.4$, the second has a drag coefficient $K=1000$ and an initial dust ratio of $0.1$. In this case, we expect the first dust species to damp the gas efficiently.

   Figure \ref{fig:dustyamp} shows the amplitude of the density perturbations (gas and dust) and velocity (gas) compared with the reference solution. Our results are in excellent agreement with the reference solution in terms of amplitude, period and damping rate. As expected the damping of the gas is significant. This emphasizes the fundamental role of the cumulative back-reaction on the dynamics of dust grains in presence of multiple species as the second phase  $K=1000$ could not damp the gas as efficiently (see Sect. \ref{sec:dustywave}).  
 \subsubsection{Example 3 : Disk settling}
\begin{figure*}[t]
       \centering
          \includegraphics[scale=0.7]{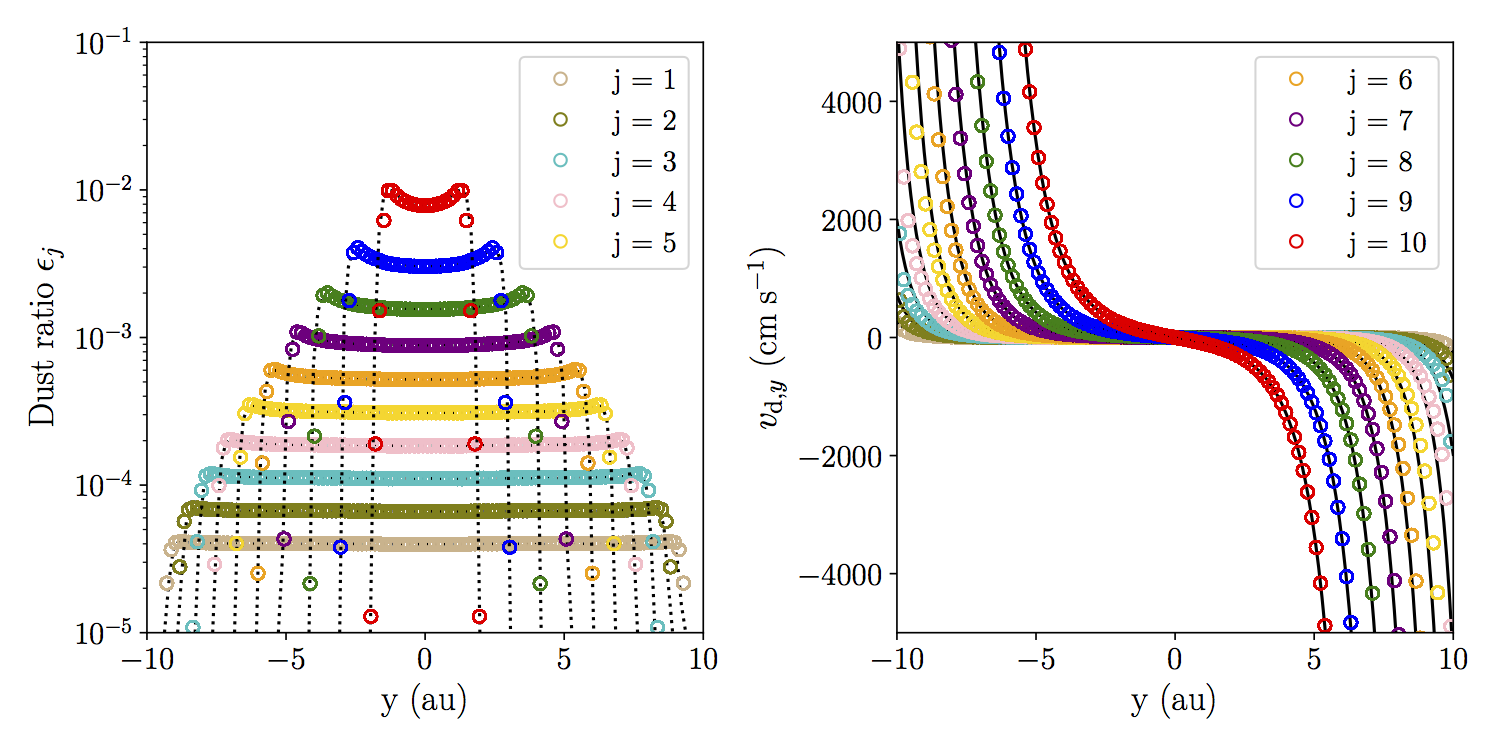}
      \caption{\textsc{multigrain} \textsc{settling} test. Dust ratios and dust velocities (circles) of the ten phases of the \textsc{settling} test after ten orbits compared with the numerical ( dotted black lines) and analytic (black lines) one-species solution.  }
       \label{fig:settlingmulti} 
\end{figure*} 

\begin{figure*}[t]
       \centering
          \includegraphics[scale=0.5]{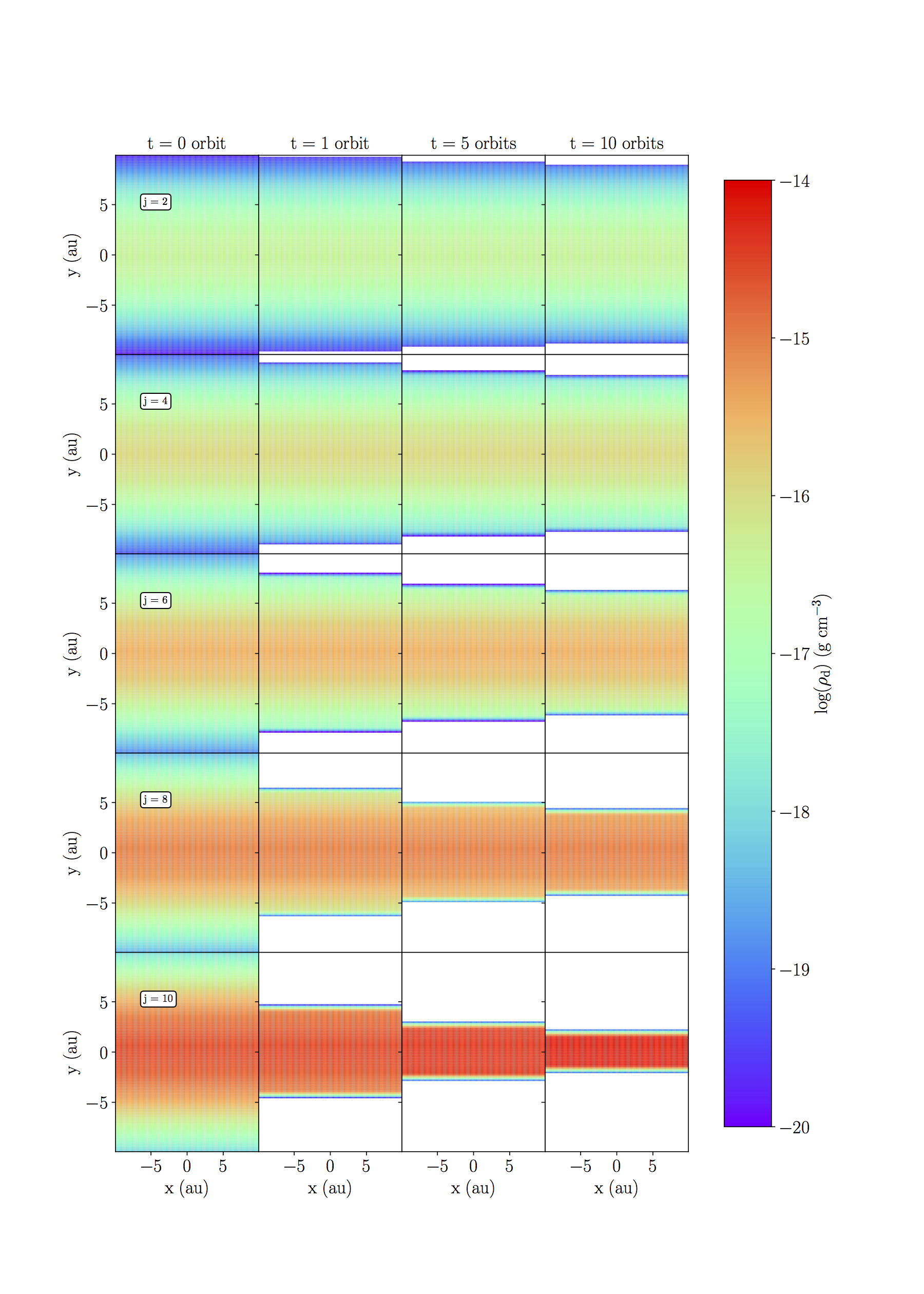}
      \caption{\textsc{multigrain} \textsc{settling} test. Dust density for the species j = 2, 4, 6, 8, and 10 in the (xy)-plane at t = 0, 1, 5, and 10 orbits.  
       \label{fig:settdens} }
\end{figure*}

\begin{figure}[t!]
       \centering
          \includegraphics[scale=0.5]{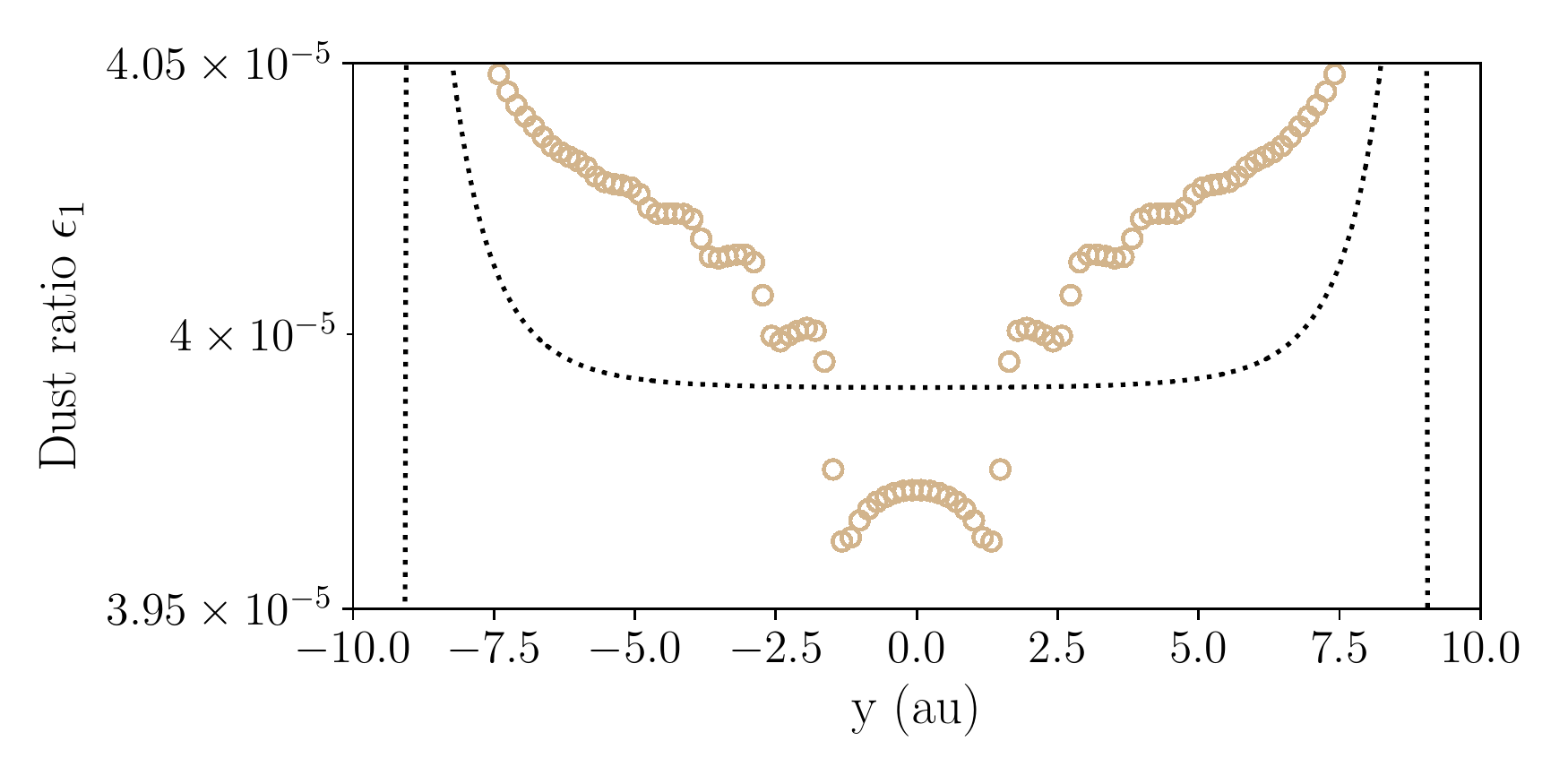}
      \caption{ \textsc{multigrain} \textsc{settling} test. Magnification of the dust front as a function of y for the $j=1$ dust species (brown circles) compared with the one-species solution (dotted line). These dust grains are dragged by the gas, which is itself submitted to the cumulative back-reaction of all the other dust species through the gas, hence the difference is the one-species solution. }
       \label{fig:zoom} 
\end{figure}

\begin{table}
       \caption{Dust distribution and stopping time at $z=0$ for the \textsc{multigrain} \textsc{settling} test}      
\label{tab:mrn}      
\centering          
\begin{tabular}{c c c c}     
\hline\hline       
                    
$j$  &  $s_{\mathrm{grain},j}$  (cm) & $\epsilon_j$ & $t_{\mathrm{s}} $ (s) at $z=0$  \\ 
\hline                    
  1 & $1.0 \times 10^{-5}$&$3.99 \times 10^{-5}$ &$40$\\
  2 & $2.78 \times 10^{-5}$&$6.65 \times 10^{-5}$ &$111$\\
  3& $7.74 \times 10^{-5}$& $1.11 \times 10^{-4}$&$310$\\
  4&$2.15 \times 10^{-4}$ & $1.85 \times 10^{-4}$&$860$\\
  5& $5.99 \times 10^{-4}$& $3.09 \times 10^{-4}$&$2396$\\
  6&$1.67 \times 10^{-3}$ &$5.15 \times 10^{-4}$&$6680$ \\
  7& $4.64 \times 10^{-3}$& $ 8.59 \times 10^{-4}$&$18560$\\
  8& $1.29 \times 10^{-2}$& $1.43 \times 10^{-3}$&$51600$\\
  9 &$3.59 \times 10^{-2}$ & $2.39 \times 10^{-3}$&$143600$\\
  10& $1.0 \times 10^{-1}$& $3.99 \times 10^{-3}$&$400000 $\\
\hline                  
\end{tabular}

\end{table}
The disk \textsc{settling} test presented in Sect. \ref{sec:sett} is nicely adapted to test our \textsc{multigrain} solver in realistic conditions. Dust grains of various sizes are present in protoplanetary disks and they experience different dynamical evolutions. To test the solver in these conditions, a settling test with ten dust species is performed with a resolution of $\ell=8$. As before, the intrinsic grain density is $3$~g~cm$^{-3}$, but every species is characterized by a proper dust ratio $\epsilon_j$ and grain size $s_{\mathrm{grain},j}$. These quantities are determined using a MRN-like distribution \footnote{$ \epsilon \left(s_{\mathrm{grain}}\right) \propto s_{\mathrm{grain}}^{4-m}$ with $  m=3.5$}, we use the same values for the minimum and maximum grain size as in \cite{2018MNRAS.476.2186H}.  The details of the size distribution are summarized in Table~\ref{tab:mrn}, which also shows the stopping time of each species in the mid-plane of the disk.

Figure \ref{fig:settlingmulti} shows the dust ratio and velocity of every species after ten orbits compared with the one-species solution, which is a good approximation as long as the cumulative back-reaction on the gas is small. Again, the solution is very well captured by our solver. 

Figure \ref{fig:settdens} shows the dust densities in the (xy)-plane. There is no dispersion of the values in the x-direction, and the initial symmetry of the problem is conserved to machine precision which originate from the Eulerian nature of the numerical scheme. As in previous studies, we see that ten orbits are enough to efficiently separate the dust phases. We note that an orbit at $50$ au is approximately $353$ years which is a few hundred times smaller than the free-fall timescale  of a typical protostellar cloud  of density $\approx 10^{-19} $~g cm$^{-3}$\citep{2000prpl.conf...59A} which is approximately $10^5-10^6$ years. An efficient settling is thus expected to happen during the collapse of this cloud, especially for large grains, e.g., $s_{\mathrm{grain}}>10^{-2}\centi\meter $ here, for which the typical settling timescale is a few orbits.

As in \cite{2018MNRAS.476.2186H}, we are interested in the effects of the interaction between different dust species. Figure \ref{fig:zoom} shows a zoom of the vertical profile of $\epsilon_1$ as a function of y. The behavior of this phase is similar to what was observed in \cite{2018MNRAS.476.2186H}. The dashed black line shows the one-species solution. As we see, for this species, the discrepancies between the one-species solution and the \textsc{multigrain} test are on the order of magnitude of the variations of $\epsilon_1$ at the dust front, which emphasizes the fundamental character equivalent stopping time in \textsc{multigrain} simulations. 

\section{Dusty protostellar collapses}
\label{sec:collapse}    
The final test of the algorithm verifies again that both gas and dust are sensitive to gravity and the 3D implementation of the solver.
\subsection{Dustycollapse}
To study the collapse of a dense core, 3D canonical test cases, introduced by \cite{1979ApJ...234..289B}, but in the presence of dust (\textsc{dustycollapse}) are performed to follow dust dynamics during the formation of the first Larson core \citep{1969MNRAS.145..271L}.
 The initial dense core, composed with gas and dust, has a mass $M_{0}$, a uniform density $\rho_0$ and a dust ratio $\epsilon_0$.  Its radius $R_0$ is set by the ratio between the thermal and gravitational energies $\alpha$
 \begin{eqnarray}
 \alpha &=& \frac{5}{2}\frac{(1-\epsilon_0) R_0 }{ \mathcal{G} M_0}\frac{k_{\mathrm{B}}T_{\mathrm{g}}}{\mu m_{\mathrm{H}}},
 \end{eqnarray}
where  $k_{\mathrm{B}}$ is the Boltzmann constant,  $ \mu $ is the mean molecular weight, $m_{\mathrm{H}}$ the hydrogen atomic mass, and $T_{\mathrm{g}}$ the initial gas temperature.

 A barotropic equation of state is used to describe the optically thick regime at which the evolution is adiabatic. It writes 
 \begin{eqnarray}
 \frac{P_{\mathrm{g}}}{(1-\epsilon)\rho} = c_{\mathrm{s}}^2  \left[1+ \left((1-\epsilon)\frac{\rho}{\rho_{\mathrm{ad}}} \right)^{\gamma-1}\right],
 \end{eqnarray}
  the evolution becoming adiabatic above the critical density $\rho_{\mathrm{ad}}=10^{-13}$~g cm$^{-3}$ \citep{1969MNRAS.145..271L}. For the sake of simplicity, this critical density is taken constant in this paper. A dependency with the dust ratio will be discussed in future works.

As in \cite{molecul}, we consider an unique grain density  $\rho_{\mathrm{grain}}=3$~g cm$^{-3}$ which is a good approximation for a combination of carbonaceous and silicate grains.  
\subsection{Setup}
All the models are set with $M_{0}=1 M_{\odot}$, $ \mu =2.31$, $ \gamma= 5/3$, and  $T_{\mathrm{g}}= 10$~K.  AMR grids of $\ell_{\mathrm{min}}=5$ and $\ell_{\mathrm{max}}=17$ are considered. With respect to the criterion given by \cite{1997ApJ...489L.179T} to avoid artificial clumps, the grid is refined to impose at least $15$ points per Jeans length. The region surrounding the core has a density that is one hundred time lower than the initial core but the same temperature.

The initial density and pressure gradients between the core and the ambient medium are numerical artifacts. To avoid unphysical variations of the dust ratio in these regions and unnecessary constraints on the timestep, the differential advection velocity $\vec{w_k}$ is set to zero outside the initial core. 

We consider the first hydrostatic core as the region where the gas density is higher than  $10^{-12.5}$~g cm$^{-3}$.
   \subsection{Validity of the diffusion approximation}
 
The diffusion approximation is valid as long as the stopping time $t_{\mathrm{s}}$ is shorter than the dynamical timescale which is, for gravitational collapses, roughly the free-fall time 

 \begin{eqnarray*}
  t_{\mathrm{ff}} \equiv \sqrt{\frac{3 \pi}{32 \mathcal{G} \rho}}.
 \end{eqnarray*}
 With an initial dust ratio of $0.01$ and a temperature of $10$~K, the Stokes number is
 \begin{eqnarray}
   \mathrm{St} \sim 0.114\left(\frac{M_0}{1 M_{\odot}}\right) \left(\frac{\rho_{\mathrm{grain}}}{3  \mathrm{\ g } \mathrm{\ cm}^{-3} }\right)  \left(\frac{s_{\mathrm{grain}}}{0.05 \mathrm{\ cm}} \right)  \left({\frac{\alpha}{0.5}}\right)^{3/2}<1.
 \end{eqnarray}
     
For a wide range of density and grain size the stopping time is short compared to the dynamical timescale. Initially, the stopping time is shorter than the free-fall timescale for grain sizes up to $ \simeq 4~\milli\meter$. However, as the Stokes number may vary, the validity of the approximation needs to be discussed during the collapse (see Sect. \ref{sec:coll}).
\subsection{Free-fall timescale for strongly coupled mixtures}

  \begin{figure}[t!]
       \centering
          \includegraphics[scale=0.57]{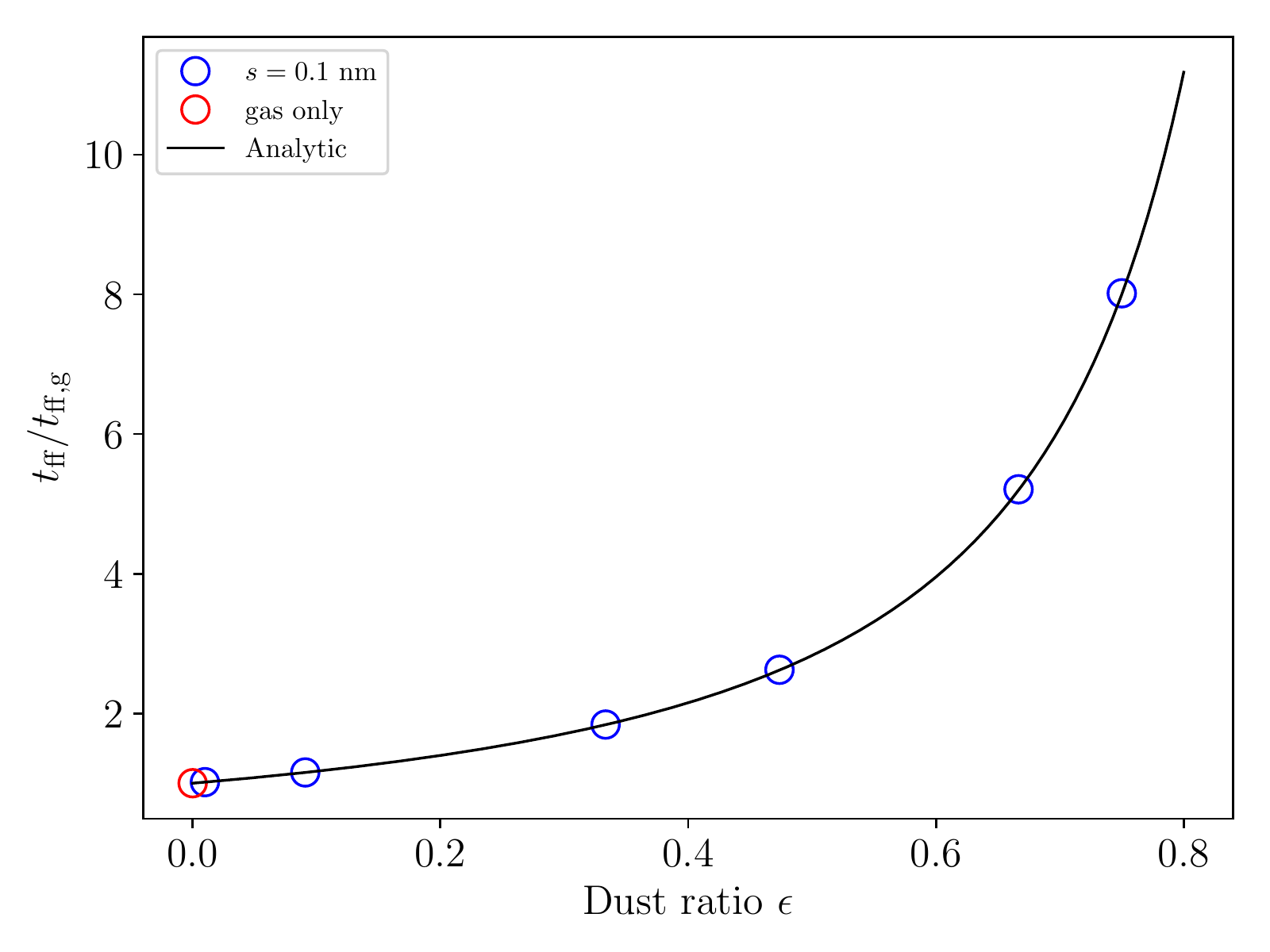}
      \caption{Ratio of the free-fall timescale of the mixture to the free-fall timescale of the gas. The red circle corresponds to the fiducial simulation with gas only, and the blue circles correspond to gas and dust mixtures for various dust ratios. The black solid line is the analytical solution.}
      \label{fig:tff}
      \end{figure}
      
 A core only composed with gas would collapse at the following free-fall timescale
 \begin{eqnarray*}
  t_{\mathrm{ff,g}} \equiv \sqrt{\frac{3 \pi}{32 \mathcal{G} \rho_{\mathrm{g}}}}.
 \end{eqnarray*}
 For a perfectly coupled gas and dust mixture ($t_{\mathrm{s}} \ll t_{\mathrm{ff}}$) it writes 
 \begin{eqnarray*}
  t_{\mathrm{ff}} = \sqrt{\frac{3 \pi}{32 \mathcal{G} \rho}}.
 \end{eqnarray*}
In the context of the Boss and Bodenheimer test, this timescale indirectly depends on the initial dust ratio
  \begin{eqnarray}
  \label{eq:tffsmall}
  t_{\mathrm{ff}} = \frac{\pi}{5\sqrt{5}}\mathcal{G}M_{0} \left( \frac{\alpha \mu m_{\mathrm{H}} }{k_{\mathrm{b}}T_{\mathrm{g}} (1-\epsilon_0)}\right)^{3/2}.
 \end{eqnarray}
 Physically, this is due to the fact that two cores with the same initial $\alpha$ but different $\epsilon_0$ have different initial radius, hence free-fall timescale. Indeed, $R_0$ increases with $\epsilon_0$ so that the gas provides the same thermal support against gravity.

To test this relation, non-rotating \textsc{dustycollapse} with various dust ratios  and the same $\alpha =0.5$ are performed. 
The condition $t_{\mathrm{s}} \ll t_{\mathrm{ff}}$ is  ensured by considering very small grains with $s =10^{-7}$~cm.  Figure \ref{fig:tff} shows the ratio of the free-fall timescale of a dusty cloud  to that of a pure-gas cloud compared with the analytical solution. We define the free-fall timescale as the time at which the peak density reaches $\rho_{\mathrm{ad}}$.  The scaling relation is very well verified. As expected, the gas, and the dust, are sensitive to gravity and fall at the mixture free-fall timescale.
 
\subsection{Core properties}
\label{sec:coll}
\begin{figure}[t!]
       \centering
          \includegraphics[scale=0.7]{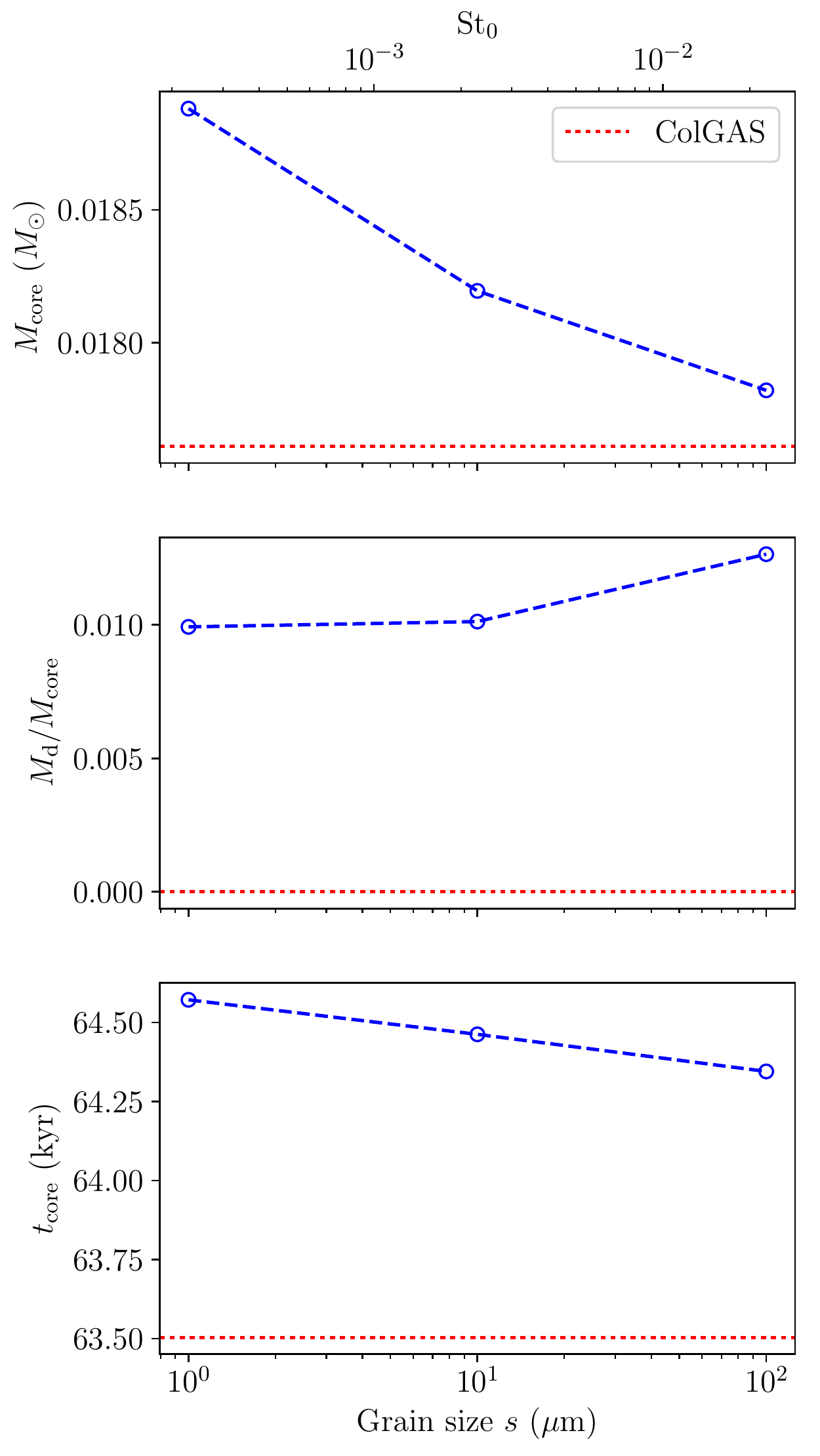}
      \caption{Properties of the first Larson core when $\rho_{\mathrm{g}}\sim 10^{-11}$~g cm$^{-3}$ for  Col1, Col10, and Col100.  The core mass, the dust mass ratio, and the formation time are shown as a function of the grain size (blue circles).  The results are compared with the fiducial ColGAS simulation (dashed red lines). The top x-axis shows the initial Stokes number in the core St$_0$.}
      \label{fig:tcore}
      \end{figure}

      \begin{figure*}[h!]
       \centering
          \includegraphics[scale=0.55]{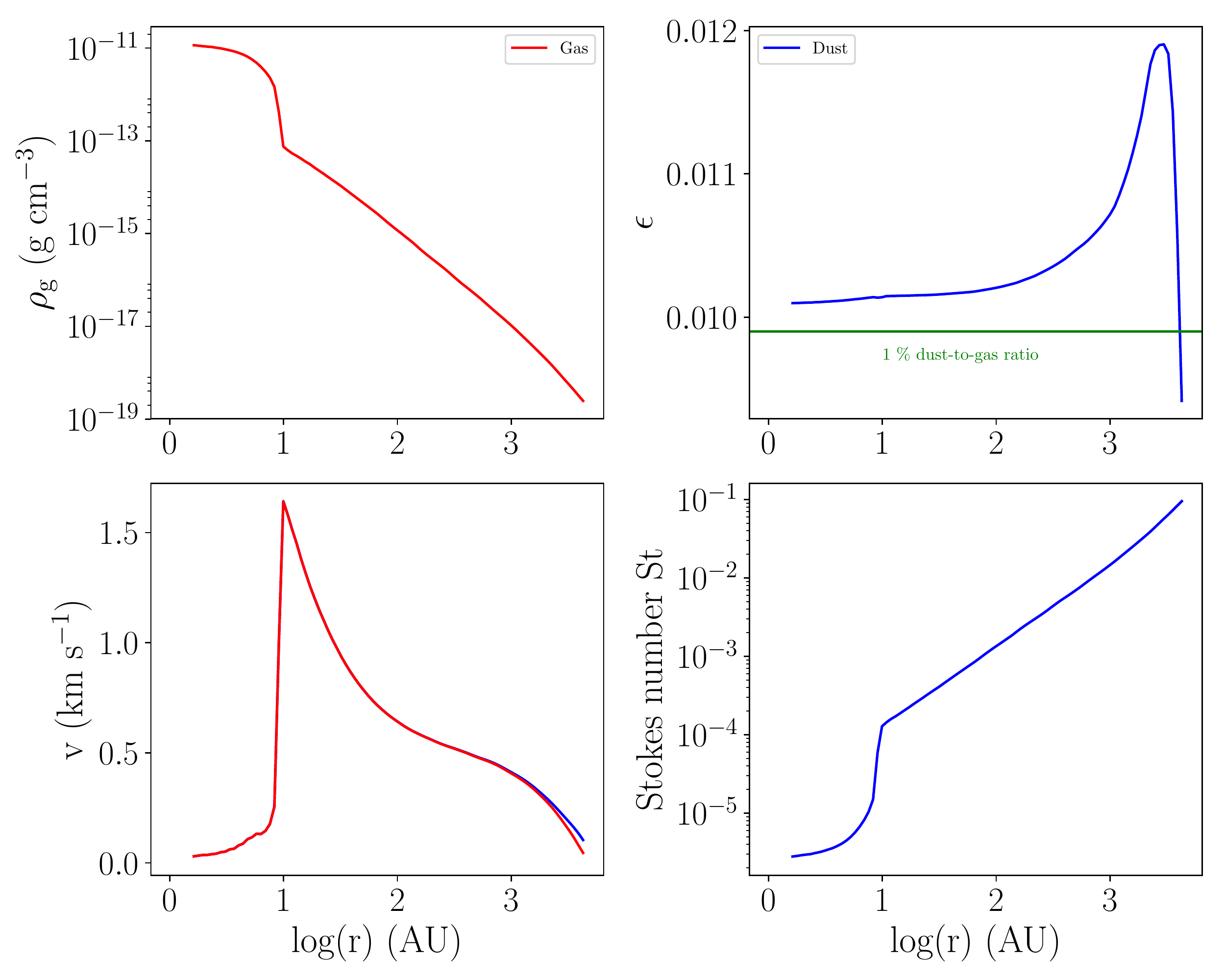}
      \caption{Radial profiles of the Col10 test. (Top Left) Gas density. (Top right) Dust ratio. (Bottom Left) Gas and dust velocities. (Bottom right) Stokes number. The horizontal green line corresponds to a dust-to-gas ratio of $1\%$}
      \label{fig:core-prof}
      \end{figure*}
   \begin{figure*}[h!]
       \centering
          \includegraphics[scale=0.55]{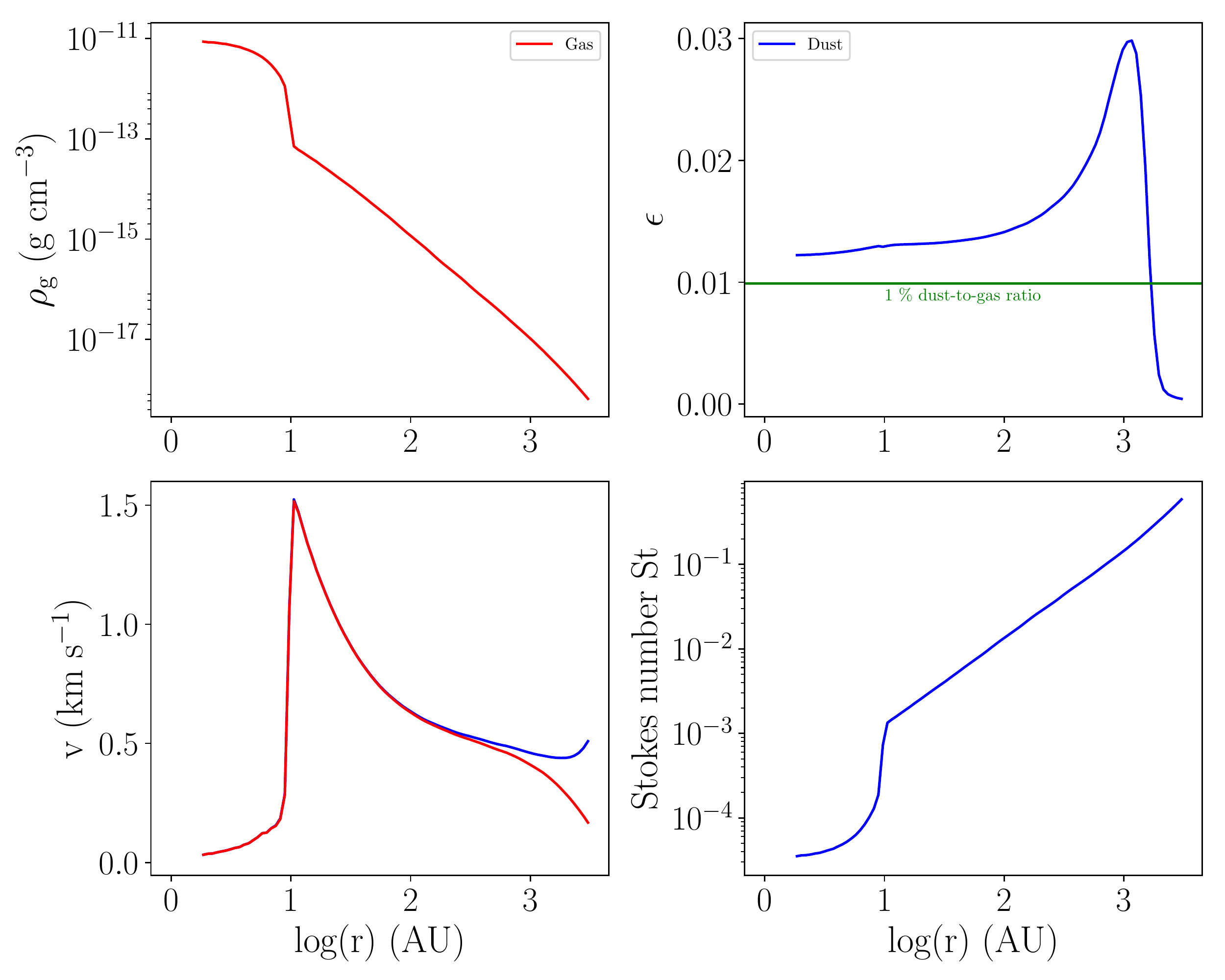}
      \caption{Same as Fig.\ref{fig:core-prof}, but for the Col100 test.}
      \label{fig:core-prof1}
      \end{figure*}
     
Non-rotating \textsc{dustycollapse} of  $\alpha=0.5$ are performed with an initial dust-to-gas ratio of $1 \%$ (corresponding to $\epsilon \sim 0.00990099$).  Single grain species of size $1~\micro\meter$ (Col1), $10~\micro\meter$ (Col10) and $100~\micro\meter$ (Col100) are considered. The results are compared, when the gas maximum density reaches $10^{-11}$~g cm$^{-3}$, with a fiducial collapse without any treatment of dust (ColGAS).

Figure \ref{fig:tcore} shows the total mass of the first hydrostatic core $M_{\mathrm{core}}$, the dust mass ratio $M_{\mathrm{d}}/ M_{\mathrm{core}}$  and its formation time $t_{\mathrm{core}}$ as a function of the grain size (bottom $x$-axis) and the initial Stokes number (top $x$-axis). The first core properties depend on the amount of dust in  the initial dense core. In the fiducial ColGAS test, the gas density is assimilated to the total density, as is often the case in the literature. This results in a lower first-core mass and a shorter formation time. We note that the barotropic equation of state implicitly  assumes a constant dust-to-gas ratio of $1 \%$,  which is slightly inconsistent. Dust-to-gas ratios might be even higher in core forming regions \citep{2016MNRAS.456.4174H,molecul} which could lead to even more important discrepancies. In terms of dynamics, small grains ($s<100~\micro\meter$) do not have a significant impact as they fall alongside the gas. However, the differential velocity between the gas and the dust is proportional to the stopping time, hence large grains ($s\ge 100~\micro\meter$) tend to fall substantially faster than the gas. Their infall provokes a slight acceleration of the first core formation as it changes the balance between the gravity and the thermal support of the gas. The mass of the core decreases as the grain size increases because the gas has less time to be accreted during the collapse. The settling of dust in the central regions of the core, however, leads to an increase in ${M_{\mathrm{d}}}/{M_{\mathrm{core}}}$. 

Figures \ref{fig:core-prof} and \ref{fig:core-prof1} show the radial profiles for the Col10 and Col100 tests, respectively, of the gas density, the  dust ratio, the gas and dust velocities, and the Stokes number. As can be seen, the strong dust depletion in the outer regions of the collapse provides the enrichment of the core. However, as the damping is more efficient in the high density regions (see the velocities $\vec{v_{\mathrm{g}}} \approx  \vec{v_{\mathrm{d}}}$), the dust ratio increases at a slower rate in the inner regions of the collapse, i.e, the first core. If the maximum increase of dust ratio in the outer regions of the collapse is about $ \sim 20 \%$ for the Col10 test, it goes up to $ \sim 300 \%$ for the Col100 test.  We see in the Col100 case that the the terminal velocity approximation probably leads to non-negligible errors as the Stokes number is close to one in the outer regions of the collapse . More accurate simulations using a future implementation of the full monofluid formalism \citep{2014MNRAS.440.2136L} should investigate these errors. 

If initial solid-body rotation is considered, it is expected that the large dust grains will settle in the mid-plane of the protoplanetary disk \citep{2017MNRAS.465.1089B}. Future works will take this into account and the presence of multiple grain species as well.

\section{Conclusions and perspectives}
\label{sec:conclusion}

We have presented an Eulerian approach to treat the dynamics of small dust grains. It is efficient in the diffusion regime and can be employed to treat multiple dust species simultaneously and efficiently. After a summary of the theory, we presented  the numerical scheme that we implemented in the AMR code {\ttfamily RAMSES}. It successfully passed the canonical validation tests for advection schemes and dust dynamics solvers, i.e, \textsc{dustyadvect}, \textsc{dustydiffuse}, \textsc{dustyshock}, \textsc{dustywave}, \textsc{settling}, and \textsc{dustycollapse}. We show that our scheme has a second-order accuracy in space  on uniform grids and intermediate between first- and second-order on AMR grids. The method also appears to efficiently treat a non-linear diffusion problem. The  \textsc{dustyshock}, \textsc{dustywave}, \textsc{settling} and \textsc{dustycollapse},  show that the waves and shock propagate at the correct velocity, and that the dust phase feels the common forces between gas and dust, e.g., gravity. The \textsc{dustydiffuse} test operated with a split dust phase shows that our method is able to efficiently treat multiple dust species simultaneously as the computation time scales in ${\sqrt{N}}$.  We emphasize the importance of the cumulative back-reaction on both the gas and the dust with the \textsc{multigrain} \textsc{dustywave} tests. We show with a \textsc{multigrain} \textsc{settling} test that our solver  efficiently treats multiple phases in a realistic astrophyiscal environment. We also observe the same interactions between the dust phases that were uncovered in \cite{2018MNRAS.476.2186H}. 

The method is applied to study dust dynamics during the formation of the first hydrostatic core. As found in \cite{2017MNRAS.465.1089B}, grains larger than  $s\simeq 100 ~\micro\meter$ can fall substantially faster than the gas. Above this size, the grains settle in the core, enriching its dust content and speeding up the collapse. We showed that the core properties are not affected very much by the dynamics of dust for small grains ($s< 100 ~\micro\meter$). However, they depend on the presence of dust itself as the thermal-to-gravitational energy ratio $\alpha$ depends on both the gas and the dust density. 

 The scheme was presented, for simplicity, in the hydrodynamical case;  however, it is implemented in the radiation non-ideal MHD version of {\ttfamily RAMSES}.  Realistic simulations of star formation would require us to consider the impact of dust on magnetic fields and  radiative transfer coupling with the mixture. In that perspective, later work will focus on taking into account the impact of the variation of the dust ratio on the opacity and the magnetic resistivities. 
   
   In the paper II we will discuss in more detail the grain dynamics during the collapse of rotating clouds with \textsc{multigrain} simulations.

\begin{acknowledgements}
  First, we sincerely thank the referee for providing extremely useful remarks and ideas which helped to significantly improve the quality of our work. 
  We acknowledge financial support from the Programme National de Physique Stellaire (PNPS) of CNRS/INSU, CEA, and CNES, France. This work was granted access to the HPC resources of CINES (Occigen) under the allocation DARI A0020407247 made by GENCI. Computations were also performed at the Common Computing Facility (CCF) of the LABEX Lyon Institute of Origins (ANR-10-LABX-0066). This work took part under the programs ISM3D and Core2disk of the PSI2 project funded by the IDEX Paris-Saclay, ANR-11-IDEX-0003-02.  This project was partly supported by the IDEXLyon project (contract n°ANR-16-IDEX-0005) under the auspices University of Lyon. Some of the \textsc{dustycollapse} plots were generated using the very efficient Osiris library developed by Neil Vaytet, Tommaso Grassi and Matthias Gonz\'alez whom we thank. In particular, we thank Matthias Gonz\'alez again for his useful advice on the paper. We also thank Daniel Price for his useful advice and Mark Hutchison for providing his solver to get the solution of the settling test. This project has received funding from the European Union's Horizon
2020 research and innovation program under the Marie
Skłodowska-Curie grant agreement No 823823.
 \end{acknowledgements}

\bibliographystyle{aa}
\bibliography{ref}

\end{document}